\documentclass[preprint,nofootinbib]{revtex4}

\usepackage{graphicx}  
\usepackage{dcolumn}   
\usepackage{bm}        
\usepackage{amssymb}
\usepackage{diagbox}
\usepackage{slashbox}
\usepackage{slashbox}

\usepackage{multirow}
\usepackage{amsmath}
\usepackage{cases}
\usepackage{placeins}
\usepackage{textcomp}
\UseRawInputEncoding

\usepackage{blindtext}
\usepackage{url}
\usepackage{graphicx}
\usepackage{url}
\usepackage{tikz}
\usepackage[bf]{caption}
\usepackage{amsmath}
\usepackage{cases} 
\usepackage{multirow}

\usepackage{epsfig}
\usepackage{makecell}
\usepackage{amsmath}
\usepackage{amsfonts}
\usepackage{amssymb}
\usepackage{mathtools}
\usepackage{url}
\usepackage{subfigure}
\usepackage{hhline}
\usepackage{color}
\usepackage{array,multirow}
\usepackage{bm}
\usepackage{makecell}
\usepackage{epsfig}
\usepackage{amsmath}
\usepackage{amsfonts}
\usepackage{amssymb}
\usepackage{url}
\usepackage{hyperref}
\usepackage{subfigure}
\usepackage{hhline}
\usepackage{color}
\usepackage{bm}
\usepackage{cancel}
\usepackage{mathtools}

\newcommand{\beqa}{\begin{eqnarray}}
\newcommand{\eeqa}{\end{eqnarray}}
\newcommand{\beq}{\begin{equation}}
\newcommand{\eeq}{\end{equation}}

\newcommand{\nn}{\nonumber}
\newcommand{\bmt}{\begin{pmatrix}}
\newcommand{\emt}{\end{pmatrix}}
\usepackage[toc,page]{appendix}
\usepackage{comment}
\usepackage{placeins}
\newcommand{\be}{\begin{equation}}
\newcommand{\ee}{\end{equation}}
\newcommand{\bea}{\begin{eqnarray}}
\newcommand{\eea}{\end{eqnarray}}

\hypersetup{
    colorlinks=true,
    citecolor=red,
    linkcolor=blue,
    filecolor=green,      
    urlcolor=magenta,
}

\begin{document}
\title{Imprints of new physics operators in the semileptonic $B \to a_1 (1260) \ell^- \bar{\nu}_\ell$  
process in SMEFT approach
}
\author{Manas Kumar Mohapatra${}^{1}$}
\email{manasmohapatra12@gmail.com}
\author{Dhiren Panda${}^{1}$}
\email{pandadhiren530@gmail.com}
\author{Rukmani Mohanta${}^{1}$}
\email{rmsp@uohyd.ac.in}
\affiliation{
\vspace{0.5cm}${}^1$School of Physics, University of Hyderabad, Hyderabad - 500046, India
}

\begin{abstract}
At present, there are several measurements of $B$ decays that exhibit discrepancies with the predictions of the Standard Model, and suggest the presence of new physics in $b\to s$ and $b \to c(u)$ quark level transitions. Motivated by the prospects of the ongoing high-luminosity $B$ factories,  we study the exclusive $B \to a_1 (1260) \ell^- \bar{\nu}_\ell$ process within the Standard Model Effective Field Theory (SMEFT) formalism, to understand the sensitivity of various new physics operators. The new physics parameters are constrained by using the experimental branching fractions of the (semi)leptonic $B \to \ell \bar{\nu}$ and $B \to (\pi, \rho, \omega) \ell \bar{\nu}$ processes (where $\ell = e, \mu, \tau$) which undergo $b \to u \ell \bar{\nu}$ quark level transitions. We then perform a comprehensive angular  analysis of the exclusive $B \to a_1 (1260) \ell^- \bar{\nu}_\ell$ process in the Standard Model and in the presence of various new physics operators. We also provide the predictions and comment on various observables, such as branching ratio, forward-backward asymmetry, and the test of lepton flavor non universality of the $B \to a_1 (1260) \ell^- \bar{\nu}_\ell$ channel.
\end{abstract}

\maketitle

\section{Introduction}
The search for physics beyond the Standard model (SM) in $B$ meson decays   has  not only attracted considerable  attention in recent times but also anticipated to remain as an active  area of research for the coming years. With thorough and careful investigation of various $B$ decays, it might be possible to get  the smoking gun signals of new physics (NP) from $B$ sector. The LHC at CERN, particularly LHCb experiment plays a pivotal role complementary to the Belle II experiment at KEK,  and the results from these two experiments impart the physicists to have a clear understanding of the nature of $b$ quark decays. In this regard, the semileptonic $B$ meson decays mediated by flavor changing neutral current ($b \to s$) and charged current ($b \to c/ u$) transitions play a crucial role in probing the sensitivity of NP beyond the SM. Though, several anomalies are observed in these decay modes,   so far none of the measurements  are statistically significant enough to provide an unambiguous signal of NP.  The future up-gradation of LHC with large data samples and improved precision can reduce the systematic errors in the existing measurements.
The most interesting observables in $b \to s \ell \ell$  transitions, which were in the limelight for quite sometime for providing the  unequivocal hints of NP, are the lepton flavor universality (LFU) violating observables  $R_K$ and $R_{K^*}$, defined as 
\bea
R_{K^{(*)}}=\frac{\mathcal{B}(B \to K^{(*)} \mu \mu)}{\mathcal{B} (B \to K^{(*)} e e)}\;.
\eea
 Recently, the updated results from LHCb \cite{LHCb:2022qnv, LHCb:2022vje} confirmed that the measured values of these observables are  consistent with their SM predictions, which are of order unity. However, there exists a variety of  other observables in $b \to s \ell \ell$ transition, e.g., the celebrated  $P_5'$ observable, branching fractions of several decay modes, which  manifest a few sigma deviations with their  SM results. 
 The LHCb~\cite{LHCb:2013ghj,LHCb:2015svh} and 
ATLAS~\cite{ATLAS:2018gqc} collaborations display a $3.3\sigma$ deviation from the SM prediction in the measurement of $P_5^{\prime}$.
The branching ratio of $B_s \to \phi\, \mu^- {\mu}^+$ decay mode indicates at $3.3\sigma$~\cite{LHCb:2021zwz,LHCb:2015wdu}
in $q^2\in[1.1,6.0]$ $\rm GeV^2$.
Moreover, the measurement of the $R_{K_S^0}$ and $R_{K^{*+}}$~\cite{LHCb:2021lvy}, also deviate from their SM predictions at $1.4\sigma$ and $1.5\sigma$, respectively. Therefore, the present scenario does not completely rule out the presence of  NP in the FCNC mediated transitions $b \to s \ell \ell$. \\ 

Analogously, for the  charged current interaction processes mediated through $b \to c \ell \nu$ transition, the lepton flavour universality violation observable is defined as $R_D=\mathcal{B}(B \to D \tau \nu)/\mathcal{B} (B \to D (e, \mu) \nu)$, which has a very precise SM prediction \cite{MILC:2015uhg, Na:2015kha, Aoki:2016frl,Bigi:2016mdz}, calculated  using the form factors by lattice QCD approach. 
In Moriond 2019, the measurement of $R_D$, announced by Belle Collaboration~\cite{Belle:2019gij}, was consistent with the previous measurement. However, its world average value  determined by HFLAV group \cite{HFLAV:2022esi} still deviates at the level of $1.4 \sigma$  from the SM prediction. Additionally, the average of various measurements of $R_{D^*}=\mathcal{B}(B \to D^* \tau \nu)/\mathcal{B} (B \to D^* (e, \mu) \nu)$ from BaBar, Belle and LHCb experimental measurements shows a tension at the level of $2.5 \sigma$~\cite{HFLAV:2022esi}. On the other hand, the measured value of the $\tau$ polarization fraction $P_{\tau}^{D^*}$ and the longitudinal polarization fraction of $D^*$ meson ~\cite{Belle:2016dyj, Belle:2017ilt, Belle:2019ewo} reported by Belle collaboration make a difference from their SM predictions at $1.6\sigma$ and $1.5\sigma$ level, respectively. Additionally, the LHCb collaboration in 2017 \cite{LHCb:2017vlu} measured the LFU violating observable $R_{J/\psi}=\mathcal{B}(B \to J/\psi \tau \nu)/\mathcal{B} (B \to J/\psi (e, \mu) \nu)$  which lies above the SM prediction at the level of $1.8 \sigma$~\cite{Dutta:2017wpq, Dutta:2017xmj}.
With these intriguing set of results in the $b\to c \ell \nu$ quark level processes, it is worth exploring   other similar deviations to understand the sensitivity of physics beyond the SM in charged current transitions, particularly those undergoing the quark level transition pertaining to $b \to u \ell \nu$ channels. Several mild discrepancies between the SM predictions and the experimental measurements have been observed in various $b \to u \ell \nu$ mediated transitions. A few processes  are measured at the $B$-factories,  although these $b \to u$ quark level transitions are CKM suppressed as compared to $b \to c$ transitions. The measured   branching fraction of the leptonic $B \to \tau \nu$ process  by Belle and BaBar ~\cite{Belle:2012egh, Belle:2015odw, BaBar:2009wmt, BaBar:2012nus}, is not in good agreement with its SM value~\cite{UTfit:2009sxh,Charles:2011va}. An upper bound on the branching fraction of $B \to \pi \tau \nu$ has been reported to be $2.5 \times 10^{-4}$  by Belle collaboration \cite{Belle:2015qal}. Additionally, the branching ratios of the exclusive $B \to \ell \nu$ and $B \to (\pi, \rho, \omega) \ell \nu$ ($\ell = \mu, e$) decays still show mild deviations from their SM results. Inspired by these set of differences between the measured values and the SM expectations, we exploit $B \to a_1 \ell \nu$ mode (where $a_1$ is an axial vector meson) in this work. The observation of the charmless hadronic $B^0 \to a_1 (1260) \pi$ decay channel by BaBar and Belle collaborations~\cite{BaBar:2004lds, BaBar:2006xju,BaBar:2006wcl,Belle:2007zox}  helps us to perform a  detailed theoretical study of the exclusive semileptonic $B \to a_1 \ell \nu$ decay process. The study of the fully differential angular distribution of this channel is quite interesting as the meson $a_1$ decays into $\rho \pi$, which provides important source of information in the SM as well as in its possible extensions. Additionally, the $\rho$ meson in the $a_1 \to \rho \pi$ process involves both the longitudinal ($\rho _{||}$) and transverse polarization ($\rho _ \perp$), which also motivates the experimental  search for this mode.  Many analyses have been performed to account for the $b \to u \ell \nu$ transitions, see, e.g., Refs.~\cite{Aliev:1999mx, Wang:2008bw, Yang:2008xw, Li:2009tx, Rajeev:2018txm, Sahoo:2017bdx, Dutta:2018vgu, Dutta:2015ueb}. Recently, the $B \to A \ell \nu$ (where $A$ is an axial vector meson) channels have been investigated in Ref.~\cite{Kang:2018jzg, Colangelo:2019axi,Colangelo:2020jmb}. A huge data samples, $(772 \pm 11) \times 10^6$ $B\bar{B}$ \cite{Belle:2018yob} and $(6.53 \pm 0.66) \times 10^6$ $B_s\bar{B}_s$ \cite{Belle:2015gho} pairs at $\Upsilon$(4S) and $\Upsilon (5S)$ resonances, respectively are collected at Belle experiment. This statistics can be increased by collecting more data by Belle II experiment. Therefore, in principle, the $B \to a_1 \ell \nu$ decay mode can be easily accessible in $B$ factory experiments in near future. \\

In this work, our aim is to explore the consequences of a model independent effective theory formalism, the so called  the Standard Model Effective Field Theory (SMEFT) approach on the exclusive semileptonic $B \to a_1 \ell \nu$ decay mode. The BSM (beyond the Standard model) physics effects can be constructed from the various SMEFT operators and can be analysed by performing fit to the associated new physics couplings. Our fit anatomy includes the experimental measurements of the branching fractions of the leptonic $B \to \ell \bar{\nu}$ and the semileptonic $B \to (\pi, \rho, \omega) \ell \bar{\nu}$ processes (where $\ell = e, \mu, \tau$). We mainly study the angular coefficient structure in the SM as well as in the presence of SMEFT NP operators. In addition, we also provide the predictions, and comment on the branching fraction and the lepton flavour universality violation observable defined by ${\cal R}_{a_1} =\mathcal{B}(B \to a_1 \tau \nu)/\mathcal{B} (B \to a_1 (e, \mu) \nu)$. \\

The remainder of the paper is structured as follows. In section II, we recapitulate the SMEFT Lagrangian with a list of most relevant dimension six operators for $b \to u \ell \nu$ transitions. In section III, we discuss the fully differential decay distributions of $B \to a_1 (\to \rho _{||} \pi)$ and $B \to a_1 (\to \rho _{\perp} \pi)$, by exploiting different set of angular coefficient functions in terms of new scalar, pseudoscalar, vector and tensor operators weighted by real couplings. A detailed analysis of angular coefficient functions are studied in presence of various NP couplings in section IV. We also focus on the branching ratio and the LFU violating observable in this section. Finally, we conclude our work in section V.

\section{Theoretical Framework}\label{Thfm}
In the absence of  direct evidence of new particles close to electroweak scale at the Large Hadron Collider, emphasis has been given to look for  indirect evidence of the existence of new  particles at scale $\Lambda_{\rm NP}$, that surpasses the electroweak scale. The SMEFT approach, one of the most efficient platforms, allows to investigate the possible NP hints in the $b$-hadron decays~\cite{Weinberg:1980wa, Coleman:1969sm, Callan:1969sn}. In this framework, the new physics effects at the scale $\Lambda _{\rm NP}$ comprises a set of higher dimensional operators that are suppressed with the NP energy scale. The form of these operators are built with the SM fields and respects the gauge symmetry $SU(3)_C \times SU(2)_L \times U(1)_Y$~\cite{Buchmuller:1985jz,Grzadkowski:2010es}. In this work, we focus on the $b \to u \ell \nu$ processes within this formalism by considering the dimension-6 operators. However, it would be very important to analyze the NP effects indirectly on the SM low energy processes. 
To understand the model-independent effects of heavy NP scale, the SMEFT effective Lagrangian at mass dimension-6 are expressed as~\cite{Greljo:2023bab} 
\bea
\mathcal{L_{\mathrm{eff}}} = \mathcal{L_{\mathrm{SM}}} + \sum_{Q_i=Q_i^\dagger} \frac{\tilde{C}_i}{\Lambda^2} Q_i + \sum_{Q_i\neq Q_i^\dagger} \left( \frac{\tilde{C}_i}{\Lambda^2} Q_i + \frac{\tilde{C}_i^\ast}{\Lambda^2} Q_i^\dagger \right) \,.
\label{Lag:SMEFT}
\eea
The relevant SMEFT dimension-six  operators contributing to $b \to u \ell \nu$ processes $Q_i$, obtained by integrating out the heavy NP particles, are given as follow
\bea 
&& Q_{lq}^{(3)} = (\bar{\ell}_p\gamma_\mu\sigma_a \ell_r)(\bar q_s \gamma^\mu\sigma^a q_t), \hspace{0.65cm}  Q_{ledq} = (\bar{\ell}_p^j e_r)(\bar d_s q_{tj}), \hspace{1.7cm} Q_{lequ}^{(1)}=(\bar{\ell}_p^j e_r)\varepsilon_{jk}(\bar q_s^k u_t),\nn\\
&&  Q_{lequ}^{(3)}=(\bar{\ell}_p^j \sigma_{\mu\nu} e_r)\varepsilon_{jk}(\bar q_s^k \sigma^{\mu\nu} u_t), \hspace{0.45cm}  Q_{\phi q}^{(3)}= (\phi ^\dagger i \overleftrightarrow{D}_\mu^a \phi) (\bar{q}_p \sigma _a \gamma ^\mu q_r), \hspace{0.4cm}  Q_{\phi ud}=(\tilde{\phi}^\dagger i D_\mu \phi) (\bar{u}_p \gamma^\mu d_r),\nn\\
&& Q_{\ell q}^{(1)}=(\bar{\ell}_p \gamma _\mu \ell _r)(\bar{q}_s \gamma^\mu q_t), \hspace{1.5cm} Q_{\phi q}^{(1)}=(\phi ^\dagger i \overleftrightarrow{D}_\mu \phi) (\bar{q}_p \gamma^\mu q_r), \hspace{0.5cm} Q_{\phi \ell}^{(3)}=(\phi ^\dagger i \overleftrightarrow{D}_\mu ^ I \phi) (\bar{\ell}_p \tau ^I \gamma^\mu \ell_r),\nn\\
\label{Op:SMEFT}
\eea
where $\overleftrightarrow{D}_\mu= D_\mu -\overleftarrow{D}_\mu$, 
$\overleftrightarrow{D}_\mu^a= \sigma^a D_\mu -\overleftarrow{D}_\mu \sigma^a$, with $\sigma^a$ as the Pauli matrices and $\epsilon_{jk}$ are the totally antisymmetric tensor in $SU(2)_L$ space. The quark $``q"$ and the lepton field $``\ell"$ shown in the above equation are doublets under $SU(2)_L$ group whereas the fermions $u, d$ and $e$  correspond to the right handed singlet fields. Out of eight relevant operators, the ($Q_{\phi q}^{(3)}$, $Q _{\phi u d}$) and $Q_ {\phi \ell} ^{(3)}$ are responsible for the modification of left and right handed $W$ boson couplings with quarks and leptons, respectively. In this analysis, we focus only on the operators $Q_{lq}^{(3)}$, $ Q_{ledq}$, $Q_{lequ}^{(1)}$ and $Q_{lequ}^{(3)}$ contributing to $b \to u \ell \nu$   transitions~\cite{Aebischer:2015fzz, Gonzalez-Alonso:2017iyc}. We neglect the operator $Q_{\phi l}^{(3)}$, which could contribute to $b\to u \ell \nu$ through a modified $W$ coupling to leptons. Regarding the leptonic vertex corrections induced by the operator $Q_{\phi l}^{(3)}$, studies discussed in~\cite{Efrati:2015eaa} have shown these corrections to be bounded at a (sub)percent level. Therefore, within the current experimental precision, the influence of this operator on the variation of $b \to u \ell \nu$ transitions is considered to be negligible.

At low energy, the most general weak effective Hamiltonian governing to $b \to u \ell \nu$ transitions is given as~\cite{Gonzalez-Alonso:2017iyc, Cirigliano:2009wk, Bhattacharya:2011qm, Aebischer:2017gaw, Jenkins:2017jig}
\bea
\mathcal{H}_\mathrm{eff} = \mathcal{H}_\mathrm{eff}^\mathrm{SM} + \frac{4 G_F}{\sqrt{2}} V_{ub} \sum_{i, \ell} C_i^{(\ell)} O_i^{(\ell)}+ \mathrm{h.c.}\,,
\label{WETHam}
\eea
where $O_i^{(\ell)}$ and $C_i ^{(\ell)}$ are the local effective operators and the Wilson coefficients encoding the NP contributions respectively. The effective mass dimension-six operators    are given as
\bea
&& O_{V_L}^{(\ell)} =
(\bar{u}_L \gamma^{\mu} b_L)(\bar{\ell}_L \gamma_\mu \nu_{\ell L}),
\hspace{1cm} O_{V_R}^{(\ell)} =
(\bar{u}_R \gamma^{\mu} b_R)(\bar{\ell}_L \gamma_\mu \nu_{\ell L}),\nn\\
&& O_{S_L}^{(\ell)} =
(\bar{u}_R b_L)(\bar{\ell}_R \nu_{\ell L}),
\hspace{1.7cm}
O_{S_R}^{(\ell)} =
(\bar{u}_L b_R)(\bar{\ell}_R \nu_{\ell L}),\nn\\
&&
O_{T}^{(\ell)} =
(\bar{u}_R \sigma^{\mu\nu} b_L)(\bar{\ell}_R \sigma_{\mu\nu} \nu_{\ell L}),
\eea 
where the $O_{V_{L,R}}^{(\ell)}$, $O_{S_{L,R}}^{(\ell)}$ and $O_{T}^{(\ell)}$ are respectively known are vector, scalar and tensor operators. Now, in terms of the dim-6 SMEFT operators, the above Wilson coefficients can be expressed as follows~\cite{Greljo:2023bab,Jenkins:2017jig}
\begin{align}
C_{V_L}^{(\ell)} &=
-\frac{V_{ud}}{V_{ub}} \frac{v^2}{\Lambda^2} \left[\tilde{C}_{\ell q}^{(3)}\right]_{\ell \ell 13},
&
C_{V_R}^{(\ell)} &=
\frac{1}{2 V_{ub}} \frac{v^2}{\Lambda^2} \left[\tilde{C}_{\phi u d}\right]_{13} \,,
\\
C_{S_L}^{(\ell)} &=
-\frac{1}{2 V_{ub}} \frac{v^2}{\Lambda^2} \left[\tilde{C}_{\ell equ}^{(1)}\right]_{\ell \ell 31}^\ast \,,
&
C_{S_R}^{(\ell)} &=
-\frac{V_{ud}}{2 V_{ub}} \frac{v^2}{\Lambda^2} \left[C_{\ell edq}\right]_{\ell \ell 31}^\ast  \,,
\\
C_{T}^{(\ell)} &=
-\frac{1}{2 V_{ub}} \frac{v^2}{\Lambda^2} \left[\tilde{C}_{\ell equ}^{(3)}\right]_{\ell \ell 31}^\ast.
\end{align}
Here,  $\Lambda$ is the NP cut-off scale expected to be at TeV range and the VEV of SM Higgs field $\phi$ is taken as $v =246$ GeV. The above NP contributions are obtained by matching the operators given in Eq (\ref{Op:SMEFT}) to the weak effective Hamiltonian in Eq.~(\ref{WETHam}) by assuming  the down  quark mass basis to be diagonal. Now, we explore how these  NP couplings affect the $b \to u \ell \nu$ decay modes explicitly in the  SMEFT formalism.
Here we discuss the exclusive $B \to \ell \nu$, and $B \to (\pi, \rho, \omega) \ell \nu $ decay modes and investigate how the  sensitivity of NP couplings arising from these processes impact  the $B \to a_1(\to \rho \pi) \ell \nu$ channel.   
Now, from the most general effective Hamiltonian governing to $b \to u \ell \nu$ transition~\cite{Bhattacharya:2011qm, Cirigliano:2009wk}, one can obtain the branching ratio of the pure leptonic $B \to \ell \nu$ decays. This braching fraction is sensitive to axial and pseudoscalar coefficients and is given as~\cite{Biancofiore:2013ki,Greljo:2023bab}
\bea
\frac{\mathcal{B} (B \to \ell \nu)}{\mathcal{B}(B \to \ell \nu) }_\mathrm{SM} = \left|1- C_A + \frac{m_B^2}{m_l (m_b+m_u)} C_P \right|^2 .
\eea
where $C_A= C_{V_R}^{(\ell)} - C_{V_L}^{(\ell)}$ and $C_P=C_{S_R}^{(\ell)} - C_{S_L}^{(\ell)}$.
Similarly, the differential decay width for the semileptonic $B \to P \ell \nu$ decay mode, sensitive to vector, scalar and tensor Wilson coefficients, is given by~\cite{Tanaka:2012nw, Sakaki:2013bfa}
\begin{equation}
\label{eq:BPlnu}
   \begin{split}
      \frac{d\Gamma(B \to P \ell \nu)/dq^2} {d\Gamma(B \to P  \ell \nu)^\mathrm{SM} / dq^2}=& 
      \left|1 + C_V^{(\ell)}\right|^2 \left[ \left( 1 + \frac{m_l^2}{2q^2} \right) H_{V,0}^{s\,2} + \frac{3}{2}\frac{m_ \ell^2}{q^2} \, H_{V,t}^{s\,2} \right] \\
      &+ \frac{3}{2} |C_S^{(\ell)}|^2 \, H_S^{s\,2} + 8|C_T^ \ell|^2 \left( 1+ \frac{2m_ \ell^2}{q^2} \right) \, H_T^{s\,2} \\
      &+ 3\mathrm{Re}[ ( 1 + C_V^{(\ell)} ) (C_S^{(\ell)^*}) ] \frac{m_ \ell}{\sqrt{q^2}} \, H_S^s H_{V,t}^s \\
      &- 12\mathrm{Re}[ ( 1 + C_V^{(\ell)}) C_T^{( \ell)*} ] \frac{m_ \ell}{\sqrt{q^2}} \, H_T^s H_{V,0}^s \biggl.\,,
   \end{split}
\end{equation}
where $C_V^{(\ell)} \equiv C_{V_L}^{(\ell)} + C_{V_R}^{(\ell)}$ and the scalar coefficient $C_S^{(\ell)} \equiv C_{S_L}^{(\ell)} + C_{S_R}^{(\ell)}$. The hadronic matrix elements $H_{(V,0), (V,t), S, T}^{s,2}$ are parameterized by the  form factors $f_{+}(q^2)$, $f_{0}(q^2)$ and $f_{T}(q^2)$.
 Now for the $B \to V \ell \nu$ (where $V$ denotes the vector meson) process, the differential decay width relative to SM is given as~\cite{Tanaka:2012nw, Sakaki:2013bfa}
 \begin{equation}
\label{eq:BVlnu}
   \begin{split}
    \frac{d\Gamma(B \to V \ell \nu)/dq^2} {d\Gamma(B \to V \ell \nu)^\mathrm{SM} / dq^2}  = & \left( |1+ C_{V_L}^{(\ell)}|^2 + |C_{V_R}^{(\ell)}|^2 \right) \left[ \left( 1 + \frac{m_l^2}{2q^2} \right) \left( H_{V,+}^2 + H_{V,-}^2 + H_{V,0}^2 \right) + \frac{3}{2}\frac{m_\ell^2}{q^2} \, H_{V,t}^2 \right] \\
      & - 2\mathrm{Re}[(1 + C_{V_L}^{(\ell)}) C_{V_R}^{(\ell)*}] \left[ \left( 1 + \frac{m_\ell^2}{2q^2} \right) \left( H_{V,0}^2 + 2 H_{V,+} H_{V,-} \right) + \frac{3}{2}\frac{m_\ell^2}{q^2} \, H_{V,t}^2 \right] \\
      &  + \frac{3}{2} |C_{S_R}^{(\ell)} - C_{S_L}^{(\ell)}|^2 \, H_S^2 + 8|C_T^\ell|^2 \left( 1+ \frac{2m_\ell^2}{q^2} \right) \left( H_{T,+}^2 + H_{T,-}^2 + H_{T,0}^2  \right) \\
      &  + 3\mathrm{Re}[ ( 1 - (C_{V_R}^{(\ell)} - C_{V_L}^{(\ell)})) (C_{S_R}^{(l)*} - C_{S_L}^{(\ell)*})  ] \frac{m_\ell}{\sqrt{q^2}} \, H_S H_{V,t} \\
      & - 12\mathrm{Re}[ (1 + C_{V_L}^{(\ell)}) C_T^{(\ell)*} ] \frac{m_\ell}{\sqrt{q^2}} \left( H_{T,0} H_{V,0} + H_{T,+} H_{V,+} - H_{T,-} H_{V,-} \right) \\
      &  + 12\mathrm{Re}[ C_{V_R}^{(\ell)} C_T^{(\ell)*} ] \frac{m_\ell}{\sqrt{q^2}} \left( H_{T,0} H_{V,0} + H_{T,+} H_{V,-} - H_{T,-} H_{V,+} \right) \,,
   \end{split}
\end{equation}
where the details of the hadronic matrix elements in terms of the form factors $A_{0,1,2}(q^2)$, $V(q^2)$ and $T_{1,2,3}(q^2)$ can be found in Ref.~\cite{Sakaki:2013bfa}.
\section{Sensitivity of new physics}
\subsection{Constraints  on the SMEFT coefficients}
In this subsection, we discuss the constraint analysis for the Wilson coefficients  of SMEFT approach. We consider the branching fractions of the $B \to \ell \nu$, $B \to \pi \ell \nu$ and $B \to V \ell \nu$ processes where $V$ represents the vector meson $V=(\omega, \rho)$, and the leptons are denoted as $ \ell =(e, \mu, \tau)$.  For  numerical computation, we use various input parameters such as mass of quarks, leptons, mesons, CKM matrix elements, Fermi coupling constant $G_F$, and the lifetime of $B$ meson etc as listed in Table \ref{tab_input}. 
\begin{table}[htbp]
\centering
\scalebox{0.9}{
\begin{tabular}{|cc|cc|cc|cc|cc|}
\hline
Parameter & Value & Parameter & Value (GeV) & Parameter & Value (GeV)  \\
\hline
\hline
$G_F$ & $1.166 \times 10^{-5}$ GeV$^{-2}$ &$m_u$ & 0.003  & $ m_B $  & $5.279$  \\
$(\tau _{B^{\pm}}, \tau _{B^0})$ & $(1.638, 1.519)\times 10^{-12}$ $s$  & $m_ c$ & $1.270$  & $m_ \pi$ & $0.139$   \\
($V_{ub}$, $V_{ud}$) & (0.0038, 0.976) & $m_b$ & 4.18  & $m_\rho$ & 0.775  \\
$f_B$ &  0.1905 & $m_e$ & 0.005  & $m_\omega$ & 0.782 \\
$\mathcal{B}_{a_1 \to \rho \pi}$ & $0.6019$ & $m_\tau$ & 1.776  & $m_{a_1}$ & 1.230  \\
\hline
\hline
\end{tabular}}
\caption{Values of the input parameters~\cite{ParticleDataGroup:2022pth} used in our analysis.}
\label{tab_input}
\end{table}

For hadronic inputs such as the form factors of $B \to \pi \ell \nu$ transition, we consider the  BCL parametrization from Ref.~\cite{Bourrely:2008za}
\bea
f_+(q^2)= \frac{1}{(1-q^2/m_{B^*}^2)}\sum_{n=0}^{N-1} a_n^+\Big[ z^n - (-1)^{n-N} \frac{n}{N}~z^N\Big]\;,~~~~
f_0(q^2)= \sum_{n=0}^{N-1} a_n^0 z^n\;.
\eea
Here the mass of $B^*$ meson is considered as  $5.325$ GeV, and $b_n^{+,0}$ are the expansion coefficients. The expansion parameter $z$ in the above equation is defined as
\bea
z\equiv z(q^2) = \frac{\sqrt{t_+-q^2}-\sqrt{t_+-t_0}}{\sqrt{t_+-q^2}+\sqrt{t_+-t_0}}\;,
\eea
where $t_+ =(m_B+m_\pi)^2$ and $t_0=(m_B+m_\pi)(\sqrt{m_B}-\sqrt{m_\pi})^2$. Also the expansion coefficients are extracted from the combined fit of the data obtained from the  $q^2$ distribution of $B \to \pi \ell \bar \nu_\ell$ and the lattice results \cite{FermilabLattice:2015mwy, FermilabLattice:2015cdh} which are given  below:
\bea
&&a_0^+=0.419 \pm 0.013,~~a_1^+=-0.495 \pm0.054,~~a_2^+=-0.43\pm0.13,~~a_3^+=0.22 \pm 0.31,\nn\\
&&a_0^0=0.510 \pm 0.019,~~a_1^0=-1.700 \pm 0.082,~~~a_2^0=-1.53\pm0.19,~~a_3^0=4.52 \pm 0.83\;.
\hspace*{1.0 true cm}
\eea
The form factor  $f_T(q^2)$ is related to $f_+^{B \to \pi} (q^2)$ through the the following relation~\cite{Colangelo:2020jmb},
\bea
f_+^{B \to \pi} (q^2)=\frac{m_B}{m_B+m_ \pi}f_T^{B \to \pi} (q^2).
\eea
We use the LCSR method~\cite{Bharucha:2015bzk} to compute the form factor for $B \to \rho$ and $B \to  \omega$ transitions. The formula is given by
\begin{equation}
F_i(q^2) = P_i(q^2) \sum_k \alpha _k^i \,\left[z(q^2)-z(0)\right]^k\,,
\label{eq:SSE}
\end{equation}
where $P_i(q^2)=1/(1-q^2/m_{R,i}^2)$. The details of the expansion coefficients $\alpha _k^i$ are provided in Ref.~\cite{Bharucha:2015bzk}. Now using the available data of these processes, we perform a naive $\chi ^2$ analysis to constraint the NP SMEFT coefficients. The relevant $\chi ^2$ is defined as
\bea
\chi^2(\tilde{C}^{\rm NP})= \sum_i  \frac{\Big ({\cal O}_i^{\rm Th}(\tilde{C}^{\rm NP}) -{\cal O}_i^{\rm Exp} \Big )^2}{(\Delta {\cal O}_i^{\rm Exp})^2+(\Delta {\cal O}_i^{\rm SM})^2},
\eea
where ${\cal O}_i ^ {\rm Th}$ and ${\cal O}_i ^ {\rm Exp}$ represent the theoretical values and the measured central values of the observables, respectively.  The denominator represents the  uncertainties associated with the SM and experimental values. From this analysis, we obtained the SMEFT new physics couplings. For completeness, we discuss for both $\mu$ and $\tau$ modes, and shown the details of the fit values in Table - \ref{tab_fits1}.

\begin{figure}[htbp]
\centering
\includegraphics[scale=0.55]{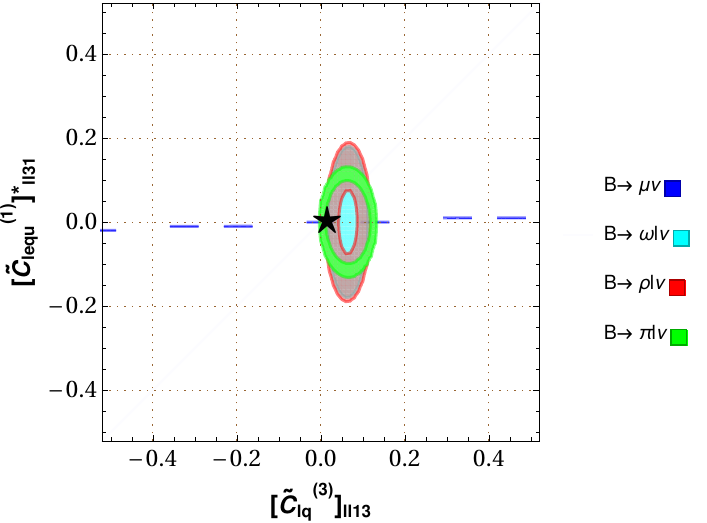}
\quad
\includegraphics[scale=0.55]{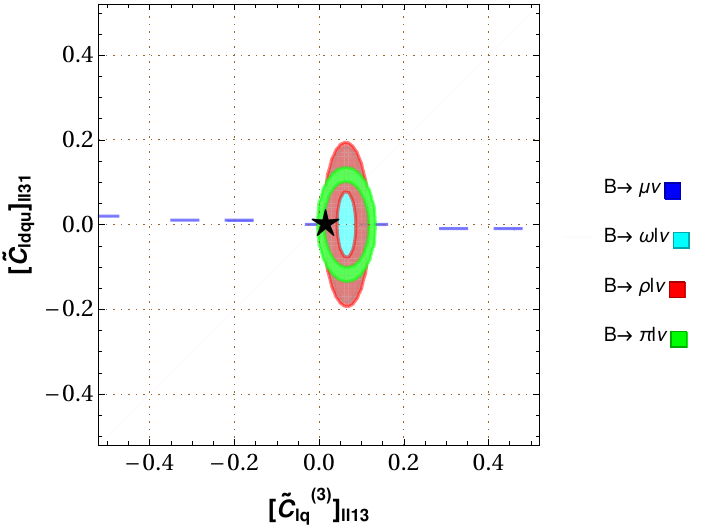}
\quad 
\includegraphics[scale=0.55]{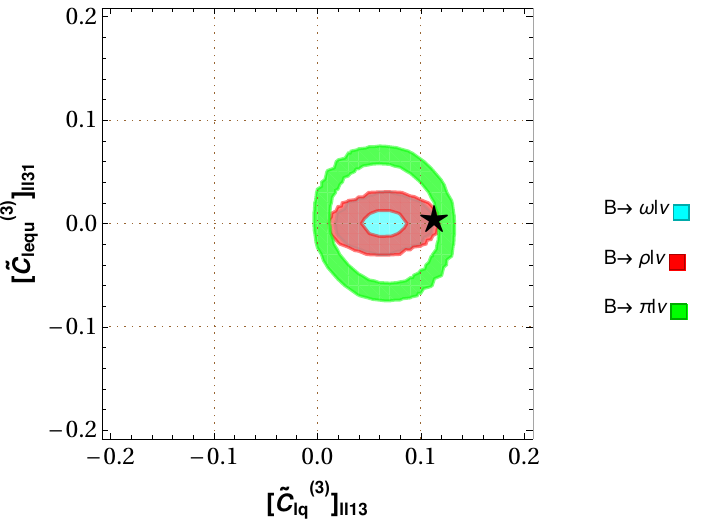}
\quad
\includegraphics[scale=0.55]{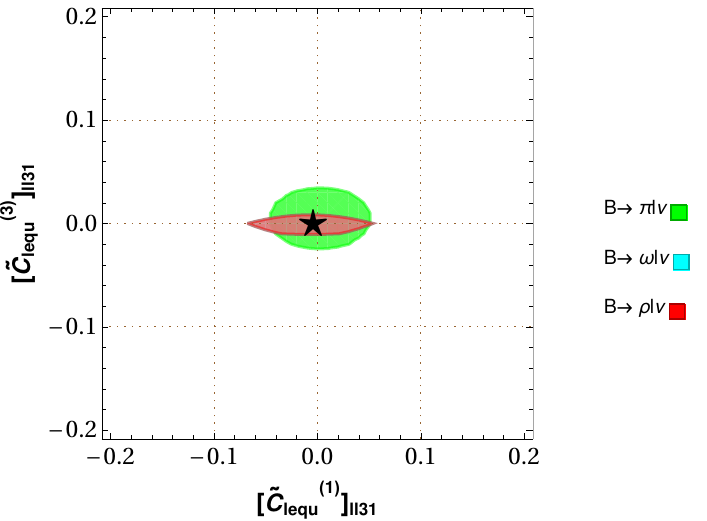}
\quad
\includegraphics[scale=0.55]{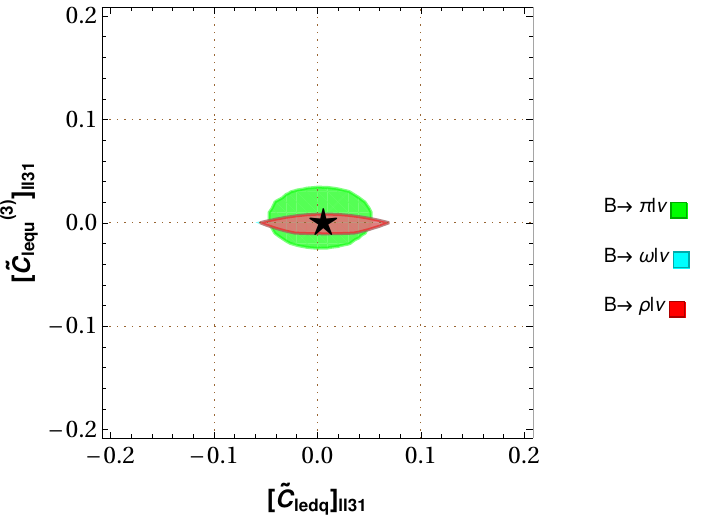}
\quad
\includegraphics[scale=0.55]{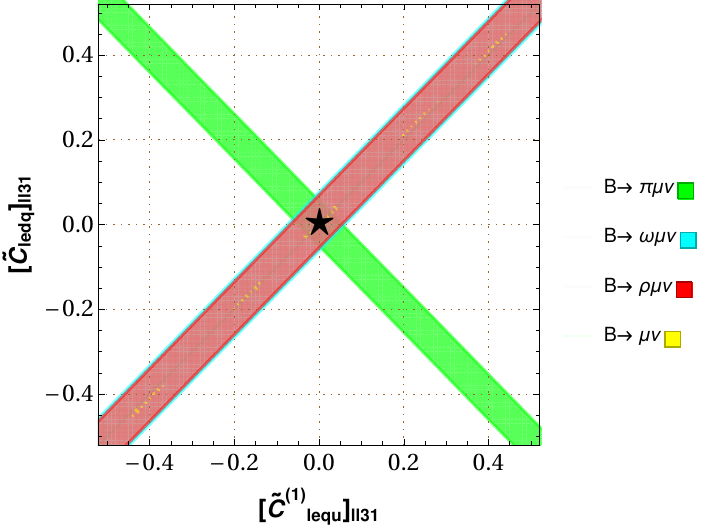}
\caption{Allowed regions for all the possible 2-D couplings from  $b \to u \mu \nu$ available data. The colors distinguish various decay modes shown in the right side of each panel. The black star represents the corresponding best-fit value.}
\label{Constraints}
\end{figure}

\begin{table}[ht]
\centering
\setlength{\tabcolsep}{7pt} 
\renewcommand{\arraystretch}{1.5} 
\begin{tabular}{|c||c||c||c|c||c|c|}
\hline
SMEFT couplings & \multicolumn{1}{c||}{Best fit ($\mu$ mode)} & \multicolumn{1}{c||}{Best fit ($\tau$ mode)}  \\
\hline
\multirow{1}{*}{$[\tilde{C}_{\ell q}^{(3)}]_{\ell \ell 13}$} & 0.013 & 0.163  \\                           
\hline
$[\tilde{C}_{\ell e q u}^{(3)}]_{\ell \ell 31}$ & -0.0008 & -0.043 \\
\hline
$[\tilde{C}_{\ell e q u}^{(1)}]_{\ell \ell 31}$ & -0.004 & -0.080 \\
\hline
$[\tilde{C}_{\ell e dq}]_{\ell \ell 31}$ & 0.005 & -0.015 \\
\hline
\hline
$([\tilde{C}_{\ell q}^{(3)}]_{\ell \ell 13}, [\tilde{C}_{\ell e q u}^{(1)}]_{\ell \ell 31})$ & (0.016, 0.001) & (-0.015, -0.080) \\
\hline
$([\tilde{C}_{\ell q}^{(3)}]_{\ell \ell 13}, [\tilde{C}_{\ell e dq}]_{\ell \ell 31})$ & (0.015, 0.004) & (-0.050,0.100) \\
\hline
$([\tilde{C}_{\ell q}^{(3)}]_{\ell \ell 13}, [\tilde{C}_{\ell e qu}^{(3)}]_{\ell \ell 31})$ & (0.113, 0.003) & (-0.030, 0.150) \\
\hline
$([\tilde{C}_{\ell e qu}^{(1)}]_{\ell \ell 31}, [\tilde{C}_{\ell e qu}^{(3)}]_{\ell \ell 31})$ & (-0.004, -0.001) & (0.060, -0.050) \\
\hline
$([\tilde{C}_{\ell e dq}]_{\ell \ell 31}, [\tilde{C}_{\ell e qu}^{(3)}]_{\ell \ell 31})$ & (0.006, -0.001) & (0.113, -0.045) \\
\hline
$([\tilde{C}_{\ell e qu}^{(1)}]_{\ell \ell 31}, [\tilde{C}_{\ell e dq}]_{\ell \ell 31})$ & (0.002, 0.0015) & (-0.050,-0.055) \\
\hline
\end{tabular}
\caption{Best fit values of SMEFT coefficients obtained from $B \to \ell \nu$ and $B \to (\pi, V) \ell \nu$ ($V = \rho, \omega$) decay channels.}
\label{tab_fits1}
\end{table}
By performing tree-level matching at the scale $\mu_\text{EW}$, we can determine the values of the Wilson coefficients for the weak effective theory (WET) operators. These coefficients can then be evolved down to the scale $\mu_b$. Numerically, the following relations hold, assuming for simplicity that all Wilson coefficients are real:
\bea
\tilde{C}_{V_L}(\mu_b) & =&-1.503\left[ \tilde{C}_{lq}^{(3)}\right]_{\ell \ell 13}(\Lambda),\nn\\
\tilde{C}_{S_L}(\mu_b) & =&-1.257\left[ \tilde{C}^{(1)}_{lequ}\right]_{\ell \ell 31}(\Lambda),\nn\\
\tilde{C}_{S_R}(\mu_b) & =&-1.254\left[ \tilde{C}_{ledq}\right]_{\ell \ell 31}(\Lambda),\nn\\
\tilde{C}_{T}(\mu_b) & =&-0.6059\left[ \tilde{C}^{(3)}_{lequ}\right]_{\ell \ell 31}(\Lambda),
\eea
where $\mu_b=4.18~\text{GeV}$ and $\Lambda=1~\text{TeV}$.
We also provide the constraint regions of all NP coefficients with the benchmark point as the $\chi ^2$ best-fit value in Fig.~\ref{Constraints}. Now, we discuss the comprehensive analysis of the angular coefficients of $B \to a_1 \ell \nu$  process.
\begin{figure}[htbp]
\centering
\includegraphics[scale=0.55]{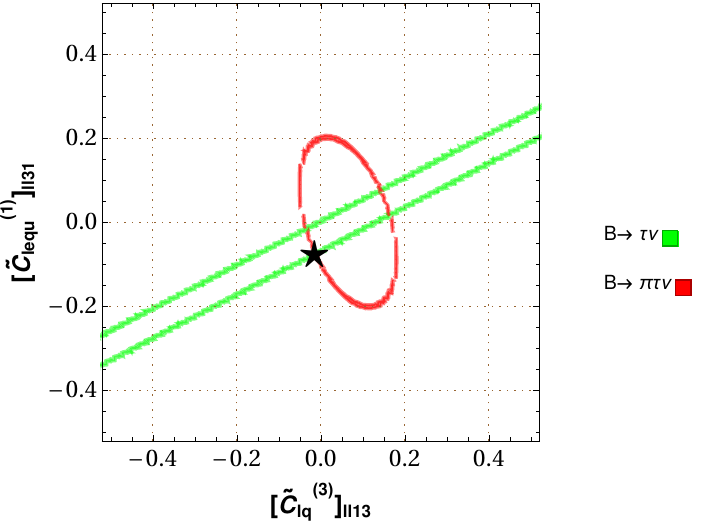}
\quad
\includegraphics[scale=0.55]{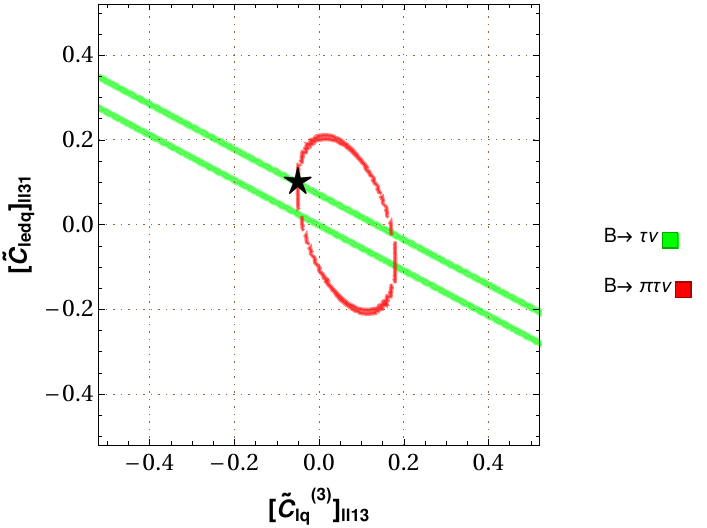}
\quad
\includegraphics[scale=0.55]{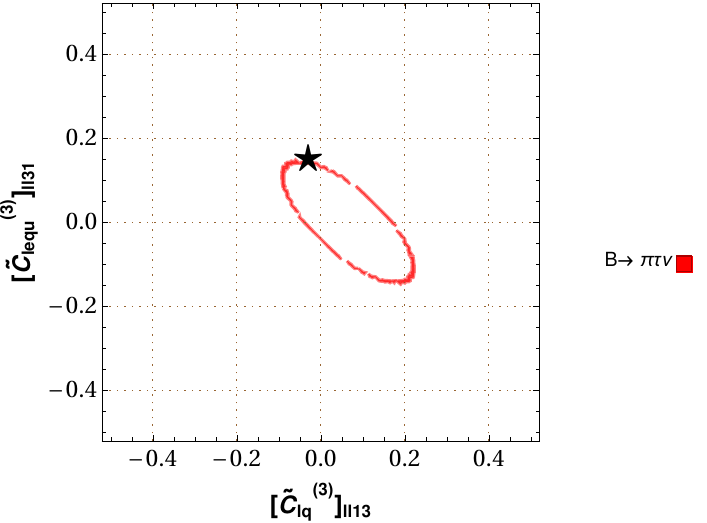}
\quad
\includegraphics[scale=0.55]{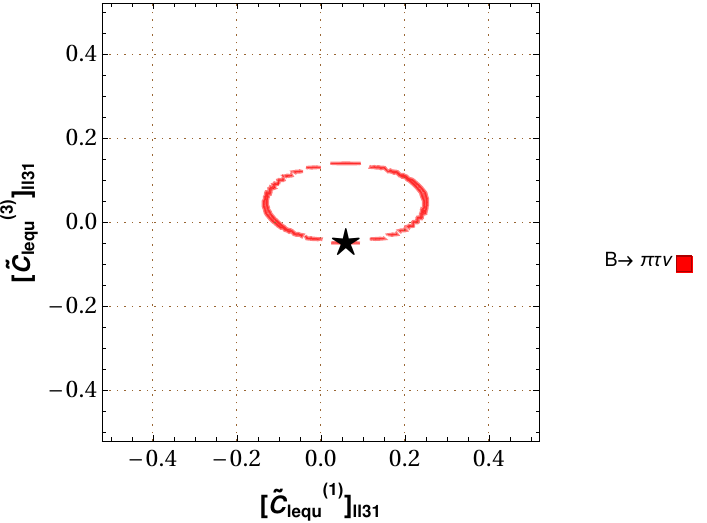}
\quad
\includegraphics[scale=0.55]{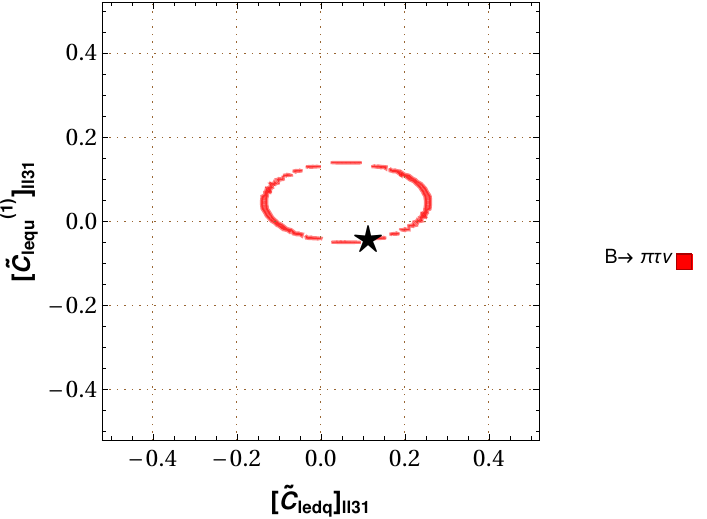}
\quad
\includegraphics[scale=0.55]{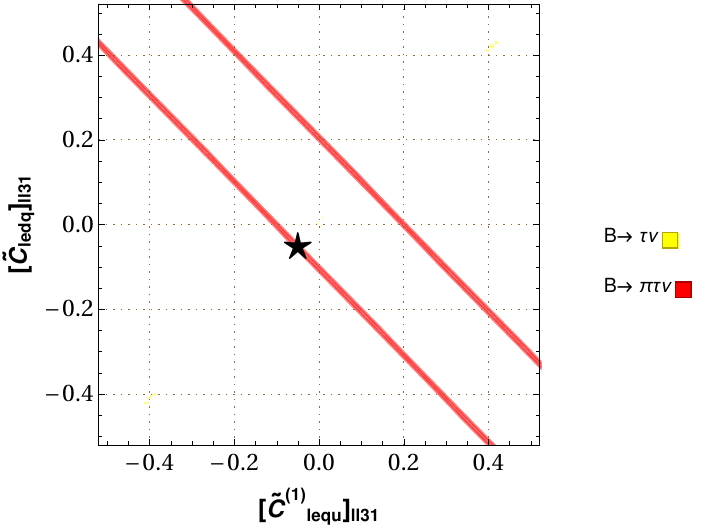}
\caption{Similar as Fig.~\ref{Constraints} for $b \to u \tau \nu$ channels.}
\label{Constraints1}
\end{figure}
\subsection{Differential Angular Analysis of $B \to a_1 (1260) \ell \nu$ Process}
It is quite interesting to visualize the $a_1 (\rho \pi)$ channel in the exclusive $B \to a_1 (1260) \ell \nu$ decay mode as the $\rho$ meson is transversely and longitudinally polarized.  
 The four dimensional differential decay distribution amplitude is given as follows \cite{Colangelo:2019axi}
 \bea
\label{angulara1}
\frac{d^4 \Gamma (\bar B \to a_1( \to \rho_{\parallel(\perp)} \pi) \ell^- \bar \nu_\ell)}{dq^2 \,d\cos \theta \,d\phi \,d\cos \theta_V}  
&=&{\cal N}_{a_1}^{\parallel(\perp)}|{\vec p}_{a_1}|\left(1- \frac{ m_\ell^2}{q^2}\right)^2 \Big\{I_{1s,\parallel(\perp)}^{a_1} \,\sin^2 \theta_V+I_{1c,\parallel(\perp)}^{a_1} \,(3+\cos 2\theta_V )\nn \\ 
&+& \left(I_{2s,\parallel(\perp)}^{a_1} \,\sin^2 \theta_V+I_{2c,\parallel(\perp)}^{a_1} \,(3+\cos 2\theta_V )\right) \cos 2\theta \nn \\   
&+&I_{3,\parallel(\perp)}^{a_1} \,\sin^2 \theta_V \sin^2 \theta  \cos  2 \phi +I_{4,\parallel(\perp)}^{a_1} \sin 2 \theta_V \sin 2\theta \cos  \phi  \nn  \\ 
&+&I_{5,\parallel(\perp)}^{a_1} \, \sin 2 \theta_V  \sin \theta \cos  \phi   \\
&+&\left(I_{6s,\parallel(\perp)}^{a_1} \, \sin^2 \theta_V+I_{6c,\parallel(\perp)}^{a_1} \,(3+\cos 2\theta_V )\right)\cos \theta
 \nn \\
&+& I_{7,\parallel(\perp)}^{a_1} \sin 2 \theta_V \sin \theta \sin  \phi \Big\}  \,\,\, . \nn
\eea
The symbol $\perp$ and $\parallel$ refer to the transverse and longitudinal polarizations of $\rho$ meson.
The expressions of the coefficients ${\cal N}_{a_1}^{\parallel(\perp)}$ and $|{\vec p}_{a_1}|$ can be read as
\bea
{\cal N}_{a_1}^{\parallel(\perp)}=\displaystyle{\frac{3G_F^2 |V_{ub}|^2 {\cal B}(a_1 \to \rho_{\parallel(\perp)} \pi)}{128(2\pi)^4m_B^2}}, \hspace{1cm} |{\vec p}_{a_1}|= \sqrt{\lambda (m_B^2, m_{a_1}^2,q^2)}/2m_B\;.
\eea
The  angular coefficient functions $I^{a1}_i$    in terms of the scalar, vector and tensor couplings are given as~\cite{Colangelo:2019axi}
\bea
I_i &=& |1+\epsilon_V|^2 \,I_i^{SM}+|\epsilon_S|^2I_i^{NP,S}+|\epsilon_T|^2I_i^{NP,T}
+2 \, {\rm Re}\left[\epsilon_S(1+\epsilon_V^* )\right] I_i^{INT,S} \nn \\ &+& 2 \, {\rm Re}\left[\epsilon_T(1+\epsilon_V^* )\right] I_i^{INT,T} +
2 \, {\rm Re}\left[\epsilon_S \epsilon_T^* \right] I_i^{INT,ST}, \,\,\,\,\hspace*{3cm}( i=1,\dots 6) ,\hspace*{0.5cm}
\label{eq:Iang} \\
I_7 &=& 2 \, {\rm Im}\left[\epsilon_X(1+\epsilon_V^* )\right] I_7^{INT,X}+2 \, {\rm Im}\left[\epsilon_T(1+\epsilon_V^* )\right] I_7^{INT,T}\nn +2 \, {\rm Im}\left[\epsilon_X \epsilon_T^* \right] I_7^{INT,XT}.
\eea
Here the NP couplings $\epsilon _i$ ($i= V, S, T$) are given as
 \begin{eqnarray}
  \epsilon_V\,= C_{V_{L}},\hspace{0.5cm} \epsilon_S=C_{S_{R}}+C_{S_{L}}\hspace{0.5cm}
  \epsilon_T=C_T\,. 
 \end{eqnarray} 
 The angular coefficient functions  $I_i^{SM}$ represent the SM contributions whereas the other coefficients such as $I_i^{NP,(S, T)}$ and $I_i^{INT, (S,T, ST)}$, expressed in terms of helicity amplitudes,  are the terms corresponding to NP and interference of NP with SM contributions respectively.
The detailed expressions are collected in Tables \ref{tab:a1SM}--\ref{tab:a1perpT}  of Appendix \ref{app:coeff}, together with the relations of the helicity amplitudes to the hadronic  form factors. Here we examine the several angular coefficient functions discussed above using the SMEFT approach.
Now, the several $q^2$ dependent observables such as branching fraction, leptonic forward-backward asymmetry in the $B\to a_1 \ell \nu$ decay mode can be obtained as~\cite{Colangelo:2019axi} 
\begin{eqnarray}
&& \mathcal{B}  (q^2)= \tau _B \frac{16}{9}\pi\frac{d\Gamma _{(||,\perp)}}{dq^2} (q^2)= \tau _B \mathcal{N}_{a_1}|\vec{p}_{a_1}|{\Big(1 - \frac{m_l^2}{q^2}\Big)}^2
 {\Big(3I_{1s(\parallel,\perp)}^{a_1}+12 \, I_{1c(\parallel,\perp)}^{a_1}-I_{2s(\parallel,\perp)}^{a_1}-4I_{2c(\parallel,\perp)}^{a_1}\Big)}\,,\nonumber \\
&&\mathcal{A}_{FB}^l(q^2)= \left[\int_0^1 \, d\cos \, \theta \, \displaystyle{\frac{d^2 \Gamma}{dq^2 d\cos \, \theta}} -\int_{-1}^0 \, d\cos \, \theta \, \displaystyle{\frac{d^2 \Gamma}{dq^2 d\cos \, \theta}} \right]\Big/{\displaystyle{\frac{d \Gamma}{dq^2}}}\nonumber \\&&~~~~~~~~~~~= \frac{8\pi}{3}\, \frac{\mathcal{N}_{a_1}|\vec{P}_{a_1}| \,\  {\Big(1 - \frac{m_l^2}{q^2}\Big)}^2
   {\Big(I_{6s}+  4I_{6c}\Big)}}{d\Gamma/dq^2}\,.
\end{eqnarray}
Alongside, we also discuss the lepton non universality (LNU) observable as,
\begin{eqnarray}
 \mathcal{R}_{a_1}(q^2) = \frac{d\Gamma (B \to a_1 \tau \nu)/dq^2}{d\Gamma(B \to a_1 (e,\mu) \nu)/dq^2}. 
\end{eqnarray}
In order to study the $B \to a_1 \ell \nu$ process, we adopt the dipole paramerization  for the form factors from Ref.~\cite{Li:2009tx}, given as
\begin{eqnarray}
 F(q^2)=\frac{F(0)}{1-a(q^2/m_B^2)+b(q^2/m_B^2)^2}\;.
 \end {eqnarray}
 The values of the form factors at $q^2=0$, i.e $F(0)$, and other relevant parameters $a$ and $b$ are given in Table \ref{Tab:formfactorsBtoa1}.
\begin{table}
 \begin{center}
 \begin{tabular}{|c|c c c|c|c c c|}
\hline \hline
 $F$       & $F(0)$  & $a$ &$b$  \\
 \hline
\hline
  $A^{B a_1}$    &$0.26_{-0.05-0.01-0.03}^{+0.06+0.00+0.03}$    &$1.72_{-0.05}^{+0.05}$    &$0.66_{-0.06}^{+0.07}$  \\
 \hline
  $V_0^{B a_1}$  &$0.34_{-0.07-0.02-0.08}^{+0.07+0.01+0.08}$    &$1.73_{-0.06}^{+0.05}$    &$0.66_{-0.08}^{+0.06}$ \\
 \hline
  $V_1^{B a_1}$  &$0.43_{-0.09-0.01-0.05}^{+0.10+0.01+0.05}$    &$0.75_{-0.05}^{+0.05}$    &$-0.12_{-0.02}^{+0.05}$   \\
 \hline
   $V_2^{B a_1}$   &$0.13_{-0.03-0.01-0.00}^{+0.03+0.00+0.00}$  &$--$  &$--$ \\
 \hline
 $T_1^{B a_1}$  &$0.34_{-0.07-0.01-0.05}^{+0.08+0.00+0.05}$    &$1.69_{-0.05}^{+0.06}$    &$0.61_{-0.05}^{+0.08}$  \\
 \hline
 $T_2^{B a_1}$  &$0.34_{-0.07-0.01-0.05}^{+0.08+0.00+0.05}$    &$0.71_{-0.05}^{+0.07}$    &$-0.16_{-0.02}^{+0.03}$\\
 \hline
  $T_3^{B a_1}$ &$0.30_{-0.06-0.01-0.05}^{+0.07+0.05+0.05}$    &$1.60_{-0.05}^{+0.06}$    &$0.53_{-0.04}^{+0.06}$\\
 \hline \hline
 \end{tabular}
 \end{center}
 \caption{$B\to a_1$ form factor inputs~\cite{Li:2009tx}.}
 \label{Tab:formfactorsBtoa1}
 \end{table}

After obtaining the NP couplings from the constraint analysis, we discuss the angular coefficients of the differential 4-dimensional distribution amplitude as well as the observables of the $B \to a_1 \ell \nu$ decay mode in the SM and in the presence of new SMEFT coefficients. We discuss the transverse and longitudinal analysis of the given process as per the behavior of $\rho$ meson. 
\section{Results and Discussions}
\label{Num_analysis}
After obtaining  the values of NP effective coefficients by exploiting the $B \to \ell \bar{\nu}$ and $B \to (\pi, \rho, \omega) \ell \bar{\nu}$ processes, we investigate  the $q^2$  distribution of the differential decay rate for the $B \to a_1 \ell \nu$ decay mode along with the angular coefficient functions. In this formalism, the new physics analysis includes the scalar, vector and tensor operators, which affect the semileptonic $B$ transitions. To understand the sensitivity of NP operators, we analyze the longitudinal and transverse contributions in SM as well as in the SMEFT approach. Below, we provide a comprehensive analysis of the longitudinal and transverse angular coefficients of $B \to a_1 \ell \nu$ channel.
\subsection{Analysis of longitudinal angular coefficients of $B \to a_1 \ell \nu$ process}
In this subsection, we discuss all the longitudinal angular coefficients $I_i^{(a_1 \to \rho _ {||}\pi)}$ of the  $B \to a_1 \ell \nu$ decay mode  in the SM as well as  in  SMEFT approach. We categorize the NP analysis in two ways such as $(S + V)$ and $(S + V + T)$. The former  includes both scalar and vector contributions whereas the later one bears the scalar, vector and tensor effective coefficients. In order to visualize the  behaviour of the given decay mode  with different leptonic  final states, we probe for $\mu$ and $\tau$ modes separately, which are depicted in Fig.~\ref{Longmu} and Fig.~\ref{Longtau}, respectively. The detailed analysis is presented below.
\begin{figure}[htbp]
\centering
\includegraphics[width=5cm,height=4.0cm]{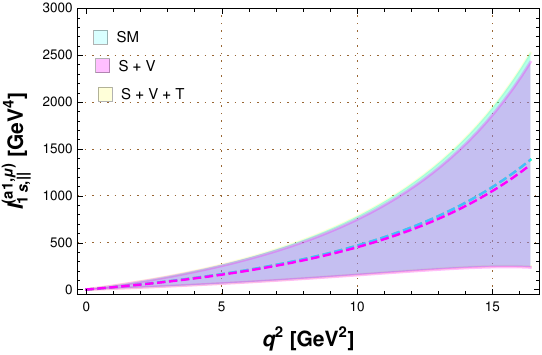}
\includegraphics[width=5cm,height=4.0cm]{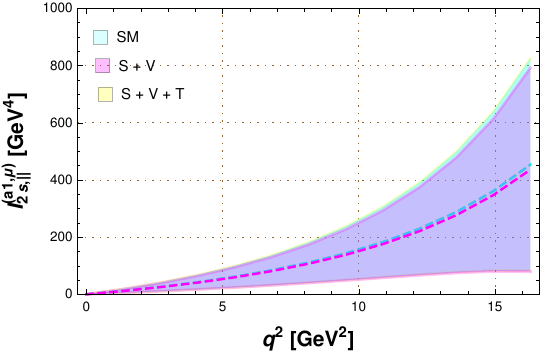}
\includegraphics[width=5cm,height=4.0cm]{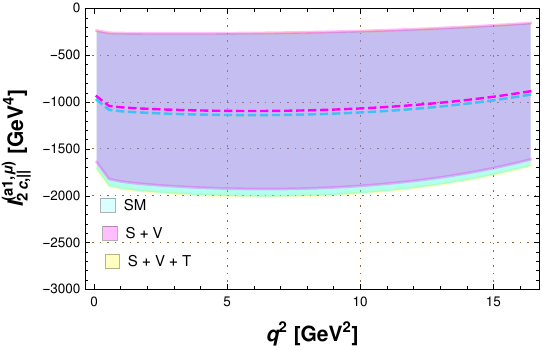}
\includegraphics[width=5cm,height=4.0cm]{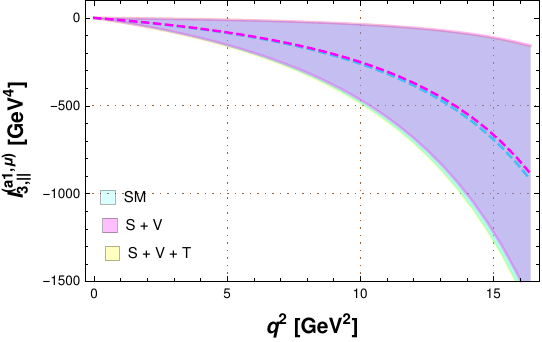}
\includegraphics[width=5cm,height=4.0cm]{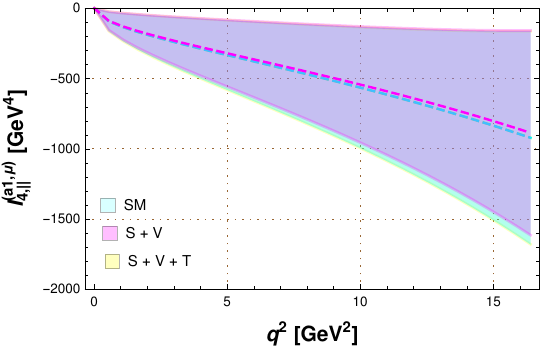}
\includegraphics[width=5cm,height=4.0cm]{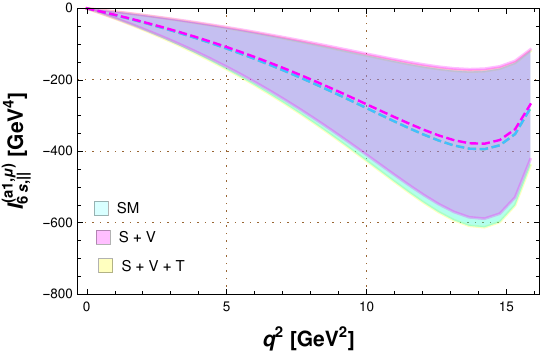}
\includegraphics[width=5cm,height=4.0cm]{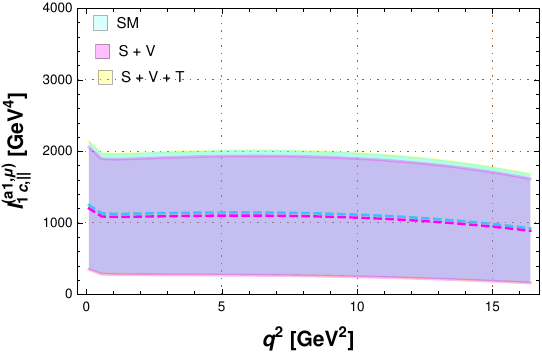}
\includegraphics[width=5cm,height=4.0cm]{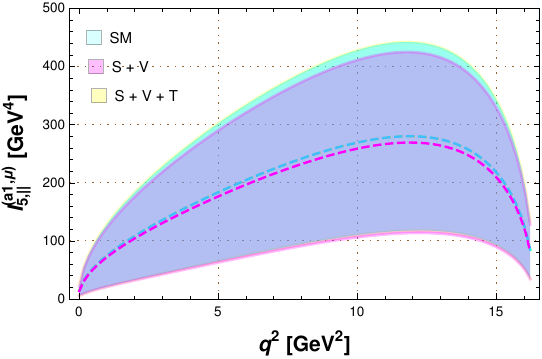}
\includegraphics[width=5cm,height=4.0cm]{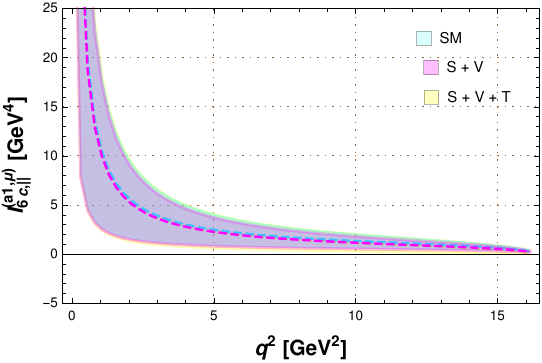}
\caption{The $q^2$ dependency of the angular coefficients of $B \to a_1 (\rho _{||} \pi) \mu \bar{\nu}$ decay mode in SM (cyan), 
$S + V$ (magenta) and $S + V + T$ (yellow).}
\label{Longmu}
\end{figure}
\subsubsection{$\mu$ mode behavior}
From the four-dimensional angular distribution of the decay width given in Eq.~(\ref{angulara1}), the complete set of longitudinal angular coefficients in SM and in the presence of various NP induced SMEFT operators can be seen in Eq.~(\ref{eq:Iang}). Now, we discuss the $q^2$ behavior of the  angular coefficients $I_{(1s, ||)}^{a_1} (q^2)$, $I_{(2s, ||)}^{a_1}(q^2)$, $I_{(2c, ||)}^{a_1}(q^2)$, $I_{(3, ||)}^{a_1}(q^2)$, $I_{(4, ||)}^{a_1}(q^2)$, $I_{(5, ||)}^{a_1}(q^2)$, $I_{(6s, ||)}^{a_1}(q^2)$, $I_{(6c, ||)}^{a_1}(q^2)$ and $I_{(1c, ||)}^{a_1}(q^2)$ (in  units of GeV$^4$), which are depicted below in Fig.~\ref{Longmu}.
As our objective is to nurture the presence of real coefficients,  the behavior of $I_7^{a_1}$ being imaginary is not considered in our analysis. The coefficient functions $I_{(1s, ||)}^{a_1} $, $I_{(2s, ||)}^{a_1}$, $I_{(2c, ||)}^{a_1}$, $I_{(3, ||)}^{a_1}$, $I_{(4, ||)}^{a_1}$, $I_{(6s, ||)}^{a_1}$, $I_{(1c, ||)}^{a_1}$, and $I_{(6s, ||)}^{a_1}$ are independent of the tensor coefficient $\tilde{C}_T$, providing only the scalar and vector effects. However, the impact of the presence of cumulative scalar and vector couplings has only slight deviation from the SM contribution, as their central values deviate marginally from the SM central value.
\subsubsection{$\tau$ mode behavior}
After the study of $\mu$ mode analysis in $B \to a_1 \ell \nu$ process, we focus on the $\tau$ lepton behavior depicted in Fig.~\ref{Longtau} as shown below. All the angular functions, both in SM, and in presence of  $(S+V)$ and $(S + V + T)$ are shown in this figure. As we mentioned before, the NP is dependent on the $\tilde{C}_{V}$ and $\tilde{C}_S$ SMEFT couplings.
\begin{figure}[htbp]
\centering
\includegraphics[width=5cm,height=4.0cm]{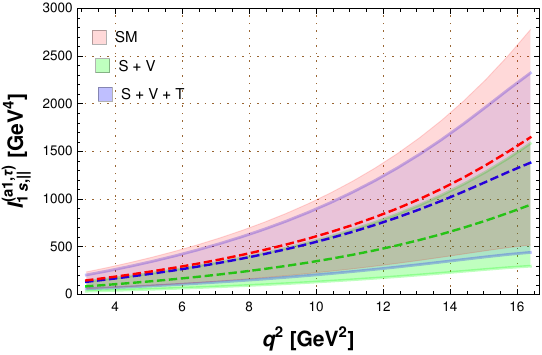}
\includegraphics[width=5cm,height=4.0cm]{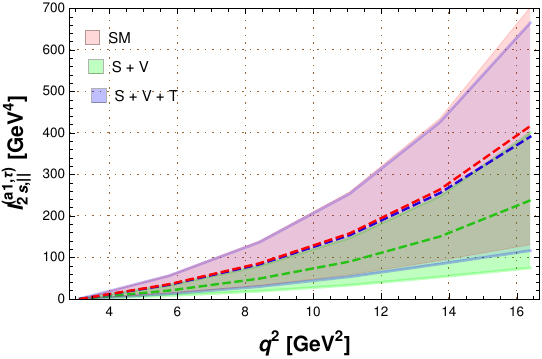}
\includegraphics[width=5cm,height=4.0cm]{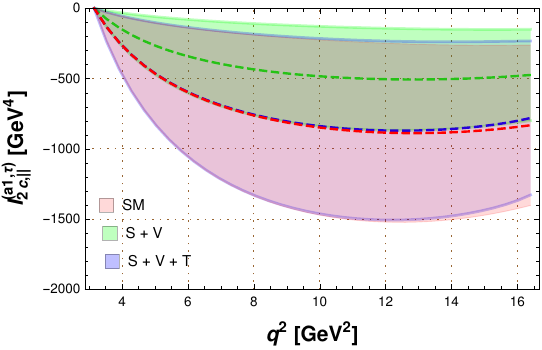}
\includegraphics[width=5cm,height=4.0cm]{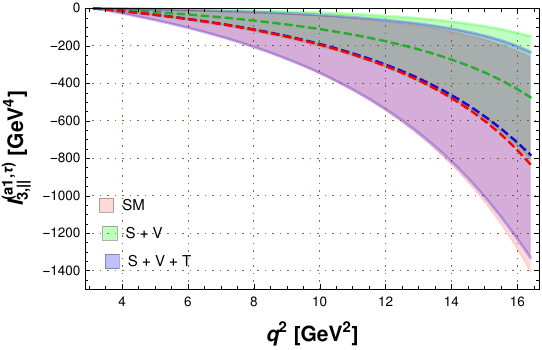}
\includegraphics[width=5cm,height=4.0cm]{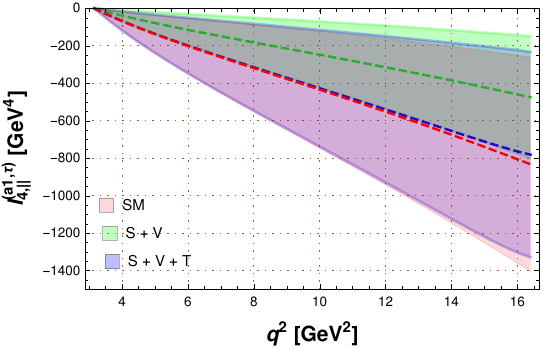}
\includegraphics[width=5cm,height=4.0cm]{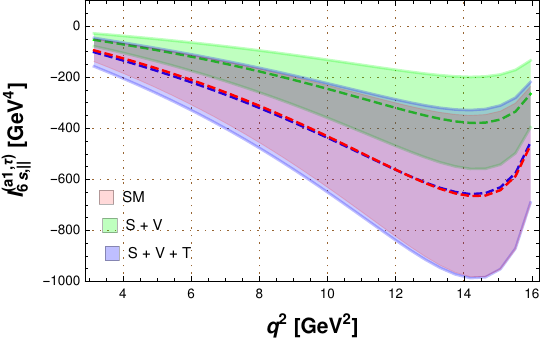}
\includegraphics[width=5cm,height=4.0cm]{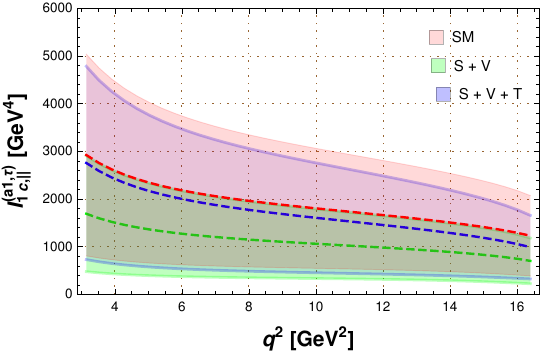}
\includegraphics[width=5cm,height=4.0cm]{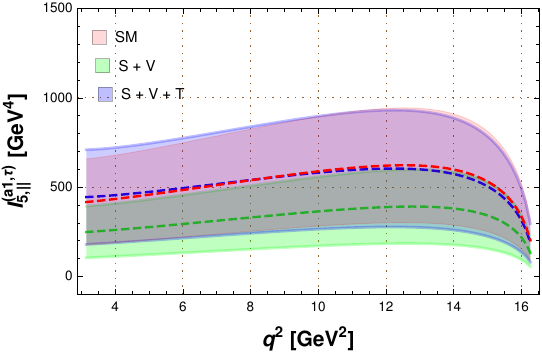}
\includegraphics[width=5cm,height=4.0cm]{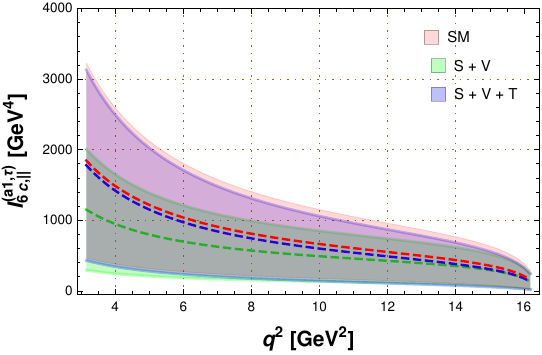}
\caption{The $q^2$ dependency of the angular coefficients of $B \to a_1 (\rho _{||} \pi) \tau \bar{\nu}$ decay mode in SM (red),  
$S + V$ (green) and $S + V + T$ (blue).}
\label{Longtau}
\end{figure}
 In Fig.~\ref{Longtau}, the red color represents the SM contribution whereas the NP contribution without (with) the tensor coupling is shown in green (blue) color. We found that the angular functions indicate remarkable deviations in the presence of the benchmark values of the NP coefficients $\tilde{C}_{V}$ and $\tilde{C}_S$. 
 However, the  inclusion of the NP tensor coefficient in addition to $\tilde{C}_S$ and $\tilde{C}_V$,  slightly shifted the values of the angular coefficients $I_{(1s, ||)}^{a_1}(q^2)$, $I_{(1c, ||)}^{a_1}(q^2)$ and $I_{(6c, ||)}^{a_1}(q^2)$  from their SM results. The 1$\sigma$ band of the NP contributions  remains more or less consistent with the SM except such angular functions. In other words, though the deviation from the SM predictions in these functions is observed in the NP scenarios, it is, however, more pronounced in the $\tau$ mode of $B \to a_1 \ell \nu$ process.  Additionally, a large hadronic uncertainty in the $\tau$ mode is obtained in the angular functions as compared to the SM. Similarly, when dealing with the $\mu$ mode, the functions exhibit a large 1$\sigma$ uncertainty, comparable to that of the SM.
\subsection{ Analysis of transverse angular coefficients in $B \to a_1 \ell \nu$ process}
\begin{figure}[htbp]
\centering
\includegraphics[width=5cm,height=4.0cm]{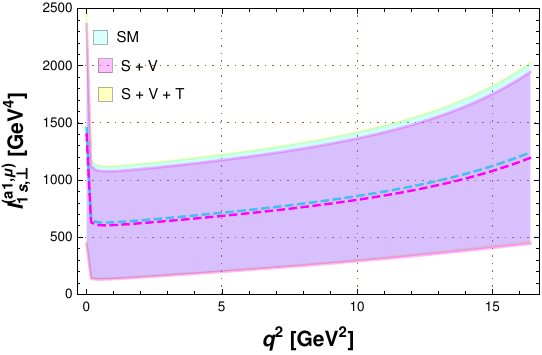}
\includegraphics[width=5cm,height=4.0cm]{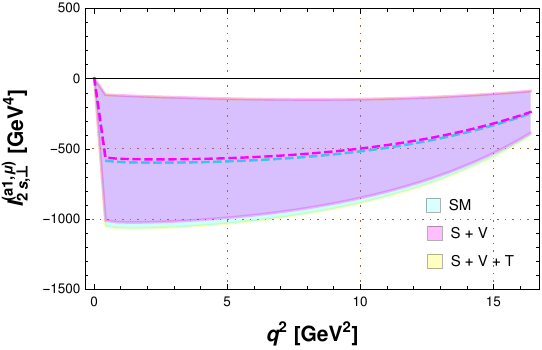}
\includegraphics[width=5cm,height=4.0cm]{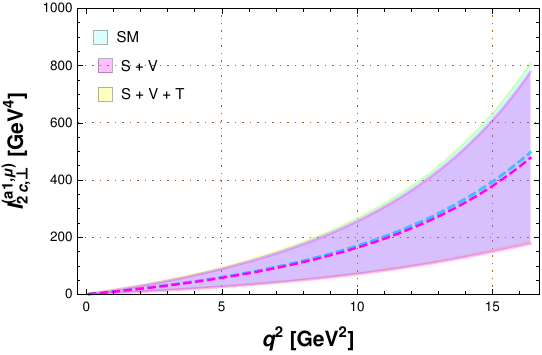}
\includegraphics[width=5cm,height=4.0cm]{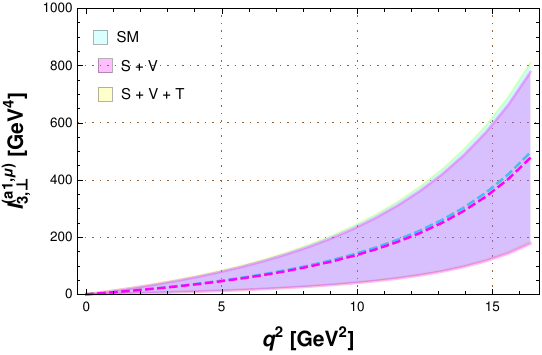}
\includegraphics[width=5cm,height=4.0cm]{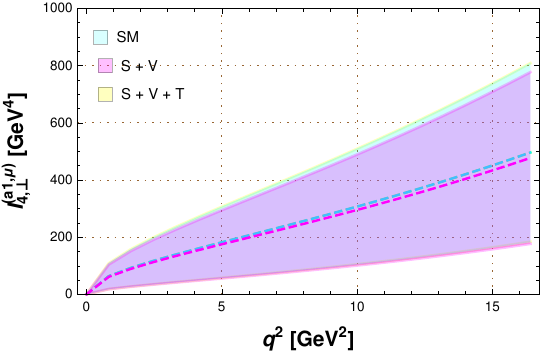}
\includegraphics[width=5cm,height=4.0cm]{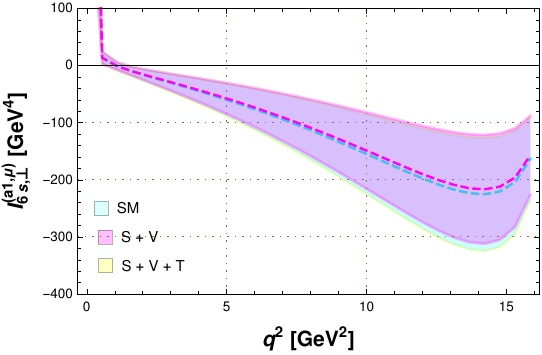}
\includegraphics[width=5cm,height=4.0cm]{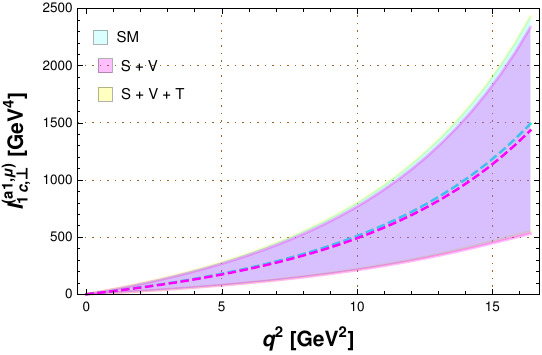}
\includegraphics[width=5cm,height=4.0cm]{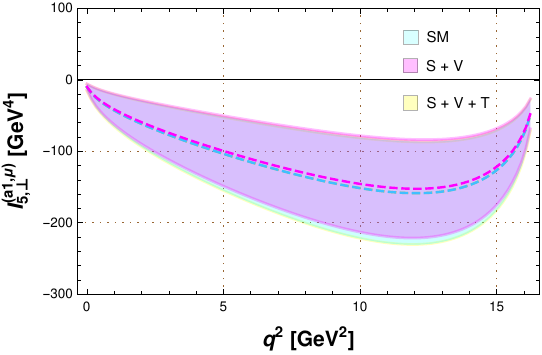}
\includegraphics[width=5cm,height=4.0cm]{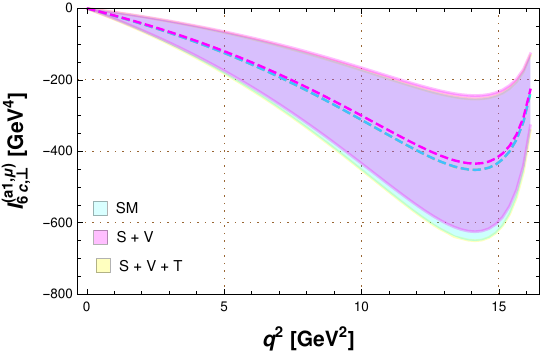}
\caption{The $q^2$ dependency of the angular coefficients of $B \to a_1 (\rho _{\perp} \pi) \mu \bar{\nu}$ decay mode in SM (cyan), 
$S + V$ (magenta) and $S + V + T$ (yellow).}
\label{Transmu}
\end{figure}

In this subsection, we discuss all the transverse angular coefficients $I_i^{(a_1 \to \rho _ {\perp}\pi)}$ of the  $B \to a_1 \ell \nu$ decay mode in SM and in the SMEFT approach. Similar to the longitudinal analysis of $B \to a_1 \ell \nu$ channel, we also study the new physics effects in two scenarios, i.e., $(S + V)$ and $(S + V + T)$.  Analogous to the longitudinal analysis, we also explore the given process for $\mu$ and $\tau$ leptonic modes. The details of the transverse analysis of $B \to a_1 \mu \nu$ and $B \to a_1 \tau \nu$ channel are depicted in Fig.~\ref{Transmu} and Fig.~\ref{Transtau}, respectively, and are discussed below.
\subsubsection{$\mu$ mode behavior}
Here, we study the muonic behavior of the  $B \to a_1 \ell \nu$ channel. The representation of the colors shown in Fig.~\ref{Transmu} are same as the longitudinal case discussed in Fig.~\ref{Longmu}.
One can observe that the coefficient functions are generally consistent and yield similar contributions in the presence of both scalar and vector NP couplings.
However, the presence of the coupling associated with the tensor operator $Q_T$ does not induce any significant deviation in the angular coefficients from their SM values. Even the central SM values are almost same as the NP contributions with tensor effective coefficient. 
\subsubsection{$\tau$ mode behavior}
\begin{figure}[htbp]
\centering
\includegraphics[width=5cm,height=4.0cm]{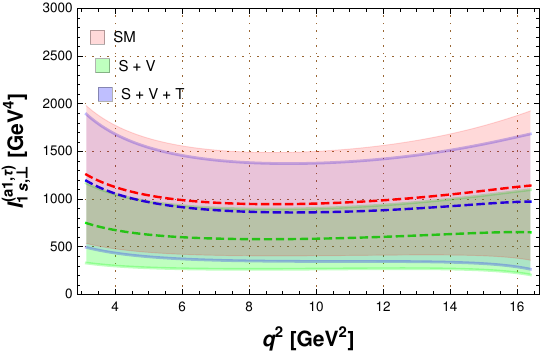}
\includegraphics[width=5cm,height=4.0cm]{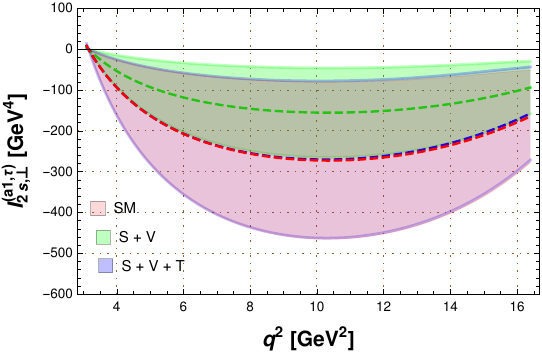}
\includegraphics[width=5cm,height=4.0cm]{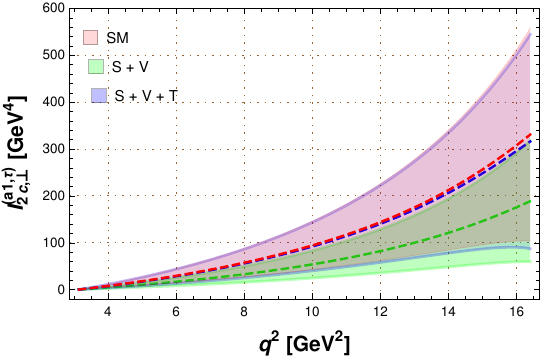}
\includegraphics[width=5cm,height=4.0cm]{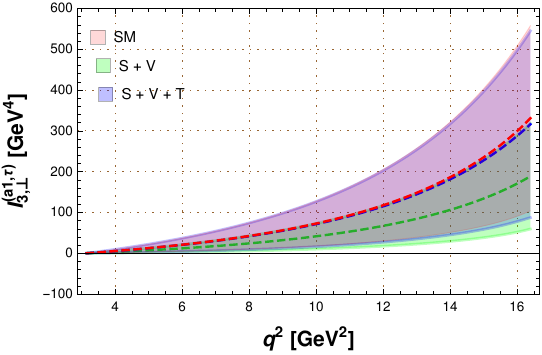}
\includegraphics[width=5cm,height=4.0cm]{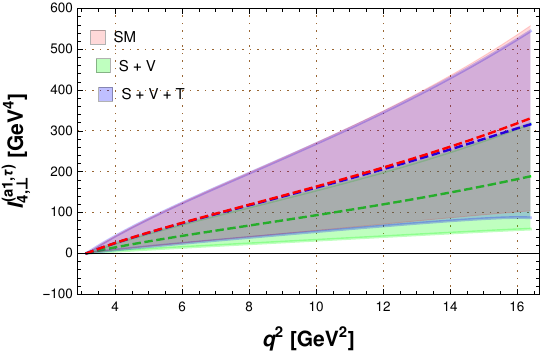}
\includegraphics[width=5cm,height=4.0cm]{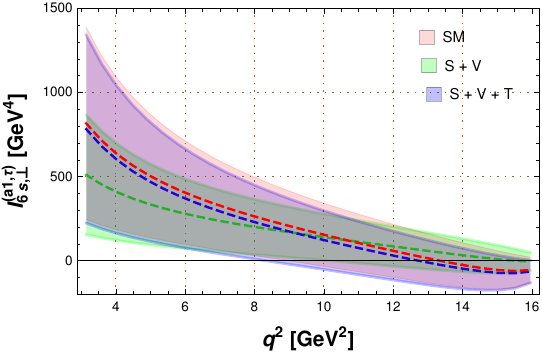}
\includegraphics[width=5cm,height=4.0cm]{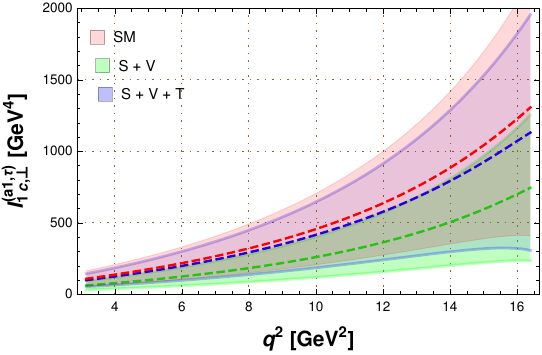}
\includegraphics[width=5cm,height=4.0cm]{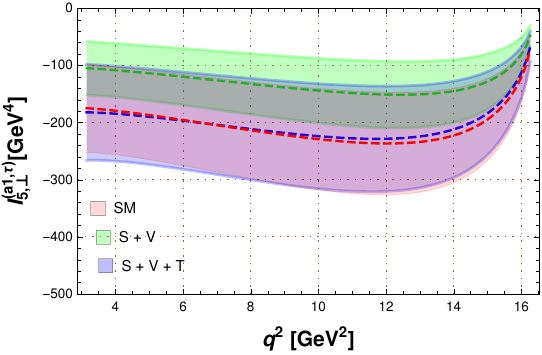}
\includegraphics[width=5cm,height=4.0cm]{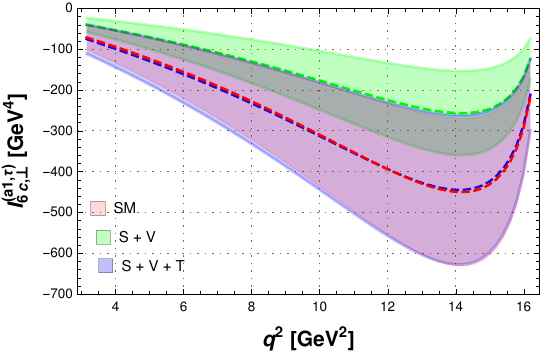}
\caption{The $q^2$ dependency of the angular coefficients of $B \to a_1 (\rho _{\perp} \pi) \tau \bar{\nu}$ decay mode in SM (red),  
$S + V$ (green) and $S + V + T $ (blue).}
\label{Transtau}
\end{figure}

\begin{figure}[htbp]
\centering
\includegraphics[scale=0.75]{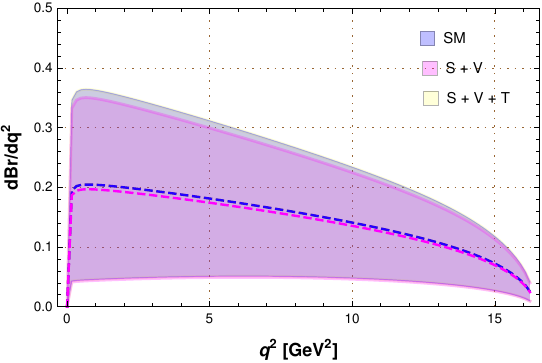}
\includegraphics[scale=0.75]{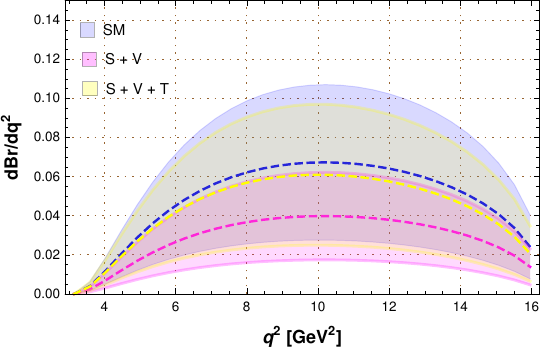}
\includegraphics[scale=0.75]{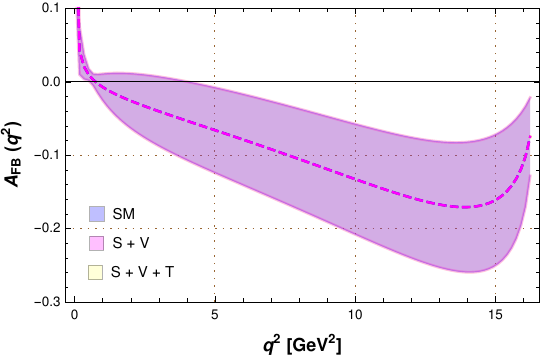}
\includegraphics[scale=0.75]{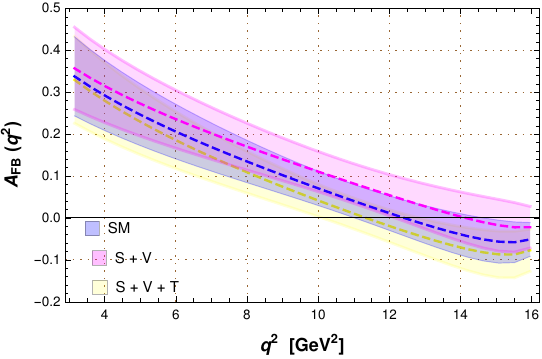}
\caption{The branching ratio (in units of $10^{-4}$) and FB asymmetry of $B \to a_1 \ell \bar{\nu}$ for $\ell = \mu$ (left) and $\ell = \tau$ (right).}
\label{BRbtoa1}
\end{figure}

In this subsection, we perform the transverse analysis of the angular coefficient functions of the $B \to a_1 \tau \nu$ decay mode in the  presence of various NP couplings. Here also the  scalar, vector and tensor operators involve in the new physics analysis. The details of the study has been shown in Fig.~\ref{Transtau}. The color representation of the plots depicted in this figure \textcolor{red}{are} same as the longitudinal analysis discussed in Fig.~\ref{Longtau}.
In the light of the NP coefficients of scalar and vector operators, the angular functions $I_{1s, \perp}^{a_1}$, $I_{2c, \perp}^{a_1}$, $I_{3, \perp}^{a_1}$, $I_{4, \perp}^{a_1}$, $I_{6s, \perp}^{a_1}$ and $I_{1c, \perp}^{a_1}$ shifted to lower values whereas the functions $I_{2s, \perp}^{a_1}$, $I_{5, \perp}^{a_1}$ and $I_{6c, \perp}^{a_1}$ get enhanced with large 1$\sigma$ uncertainties.  Now in order to have an insight of the tensor new physics coupling in addition to both $\tilde{C}_V$ and $\tilde{C}_S$, the angular coefficient functions $I_i$ are found to have  deviations remarkably. The tensor coupling even shifts the functions $I_{1s, \perp}^{a_1}$, $I_{6s, \perp}^{a_1}$ and $I_{1c, \perp}^{a_1}$ to higher values as compared to the both scalar and vector contributions. However, the angular functions $I_{2c, \perp}^{a_1}$, $I_{3, \perp}^{a_1}$ and $I_{4, \perp}^{a_1}$ shift downward and significantly deviate from the SM values. \\ 
After the analysis of the angular coefficient functions for $B \to a_1 (\to \rho _{||}\pi) \ell \nu $ and $B \to a_1 (\to \rho _{\perp}\pi) \ell \nu$ channels, we now proceed to explore other observables  of the exclusive $B \to a_1 \ell \nu$ decay mode. In particular, we will focus on the branching ratio ($\mathcal{B}$) and the LFU violating observable $\mathcal{R}_{a_1}$. The $q^2$ variation of the branching ratios are shown in Fig.~\ref{BRbtoa1}. The colors  representing different contributions are  presented in the plot legends. The left panel of the figure depicts $B \to a_1 \mu \nu$ channel. One can notice from the figure that, the branching fraction is reduced  with respect to the SM result in the presence of the operators without tensor. However, with the inclusion of tensor coupling, it remains consistent with its SM value.
On the other hand, the right panel of the Fig.~\ref{BRbtoa1} represents the branching ratio of $B \to a_1 \tau \nu$ mode. In the presence of  $(S + V)$ and $(S + V + T)$ type NP contributions, the branching ratio shows an appreciable deviation. In the lower left (right) panel, we display the variations of the forward-backward asymmetry of $B \to a_1 \mu \nu$ ($B \to a_1 \tau \nu$)  mode. It should be noted that the NP contribution to forward-backward asymmetry has a significant effect in both the $(S+V)$ and $(S+V+T)$ scenarios for $B \to a_1 \tau \nu$.
\begin{table}
 \begin{center}
\begin{tabular}{|l||*{5}{c|}}\hline
\backslashbox{Observable}{Prediction}
&\makebox[10em]{SM}&\makebox[10em]{NP with $(S + V)$}&\makebox[10em]{NP with $(S + V + T)$}
\\\hline\hline
\hspace{1cm} ${\cal B}_\mu (\times 10^{-4})$ & $(2.053 \pm 1.550)$ & $(1.974 \pm 1.518)$ & $(2.053 \pm 1.553)$\\\hline
\hspace{1cm} ${\cal B}_\tau (\times 10^{-4})$ & $(0.679 \pm 0.341)$ & $(0.394 \pm 0.216)$ & $(0.604 \pm 0.302)$\\\hline
\hspace{1cm} $A_{FB}^\mu$ & $(-0.068 \pm 0.041)$ & $(-0.068 \pm 0.040)$ & $(-0.067 \pm 0.039)$\\\hline
\hspace{1cm} $A_{FB}^\tau$ & $(0.037 \pm 0.042)$ & $(0.069 \pm 0.072)$ & $(0.028 \pm 0.037)$\\\hline
\hspace{1cm} ${\cal R}_{a_1}$ & $(0.330 \pm 0.289)$ & $(0.199 \pm 0.179)$ & $(0.294 \pm 0.257)$\\\hline
\end{tabular}
 \end{center}
 \caption{The predictions of the observables of $B \to a_1 \ell \nu$ process in SM and in SMEFT approach.}
 \label{Tab:Numresults}
 \end{table}

\begin{figure}[htbp]
\centering
\includegraphics[width=6cm,height=4.5cm]{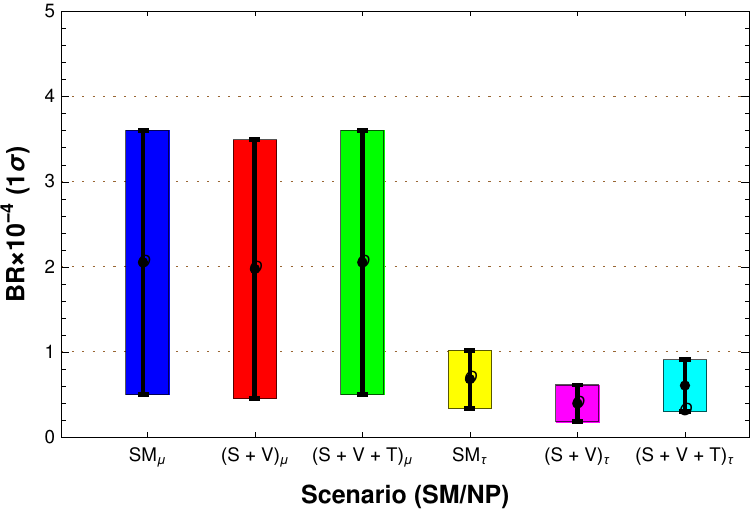}
\quad
\includegraphics[width=6cm,height=4.5cm]{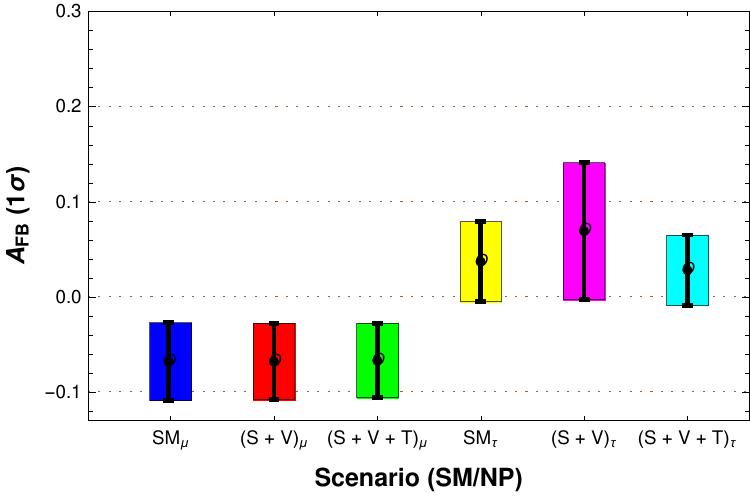}
\includegraphics[width=6cm,height=4.5cm]{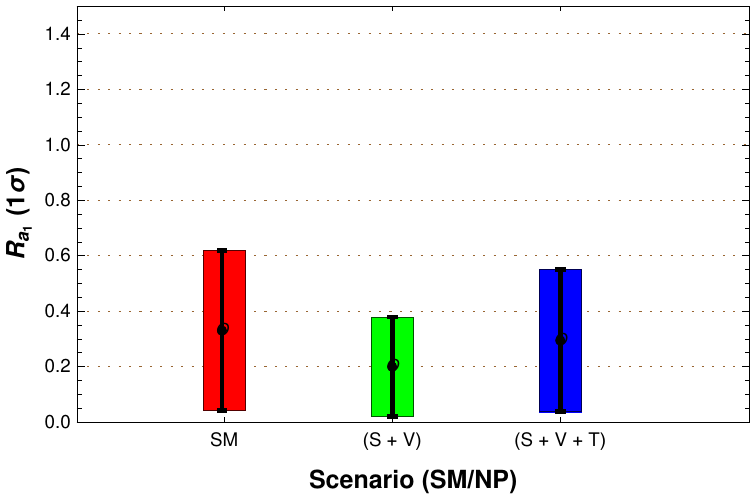}
\caption{Predictions of the ${\cal B}$, $A_{FB}$ and $R_{a_1}$ ($1\sigma$ range) of $B \to a_1 \ell \bar{\nu}$ process, respectively for the allowed solutions corresponding to $(S, V)$ and $(S, V, T)$ operators.}
\label{OBsbtoa1}
\end{figure}
The predictions of the LFU violating observable in addition to the ${\cal B}$ and $A_{FB}$ are coherently presented  in Fig.~\ref{OBsbtoa1}. For completeness, the outcomes of this investigation are summarized below. 
\begin{itemize}
\item ${\cal B}$ : The simultaneous presence of the scalar and vector operators reduce the branching ratio ($\mu$ mode) while in the presence of $(S+V+T)$ couplings, it becomes consistent with the SM prediction. On the other hand, the branching fraction ($\tau$ mode) is reduced in both of the NP scenarios, and has significant deviation from the SM.
\item $A_{FB}$ : We observe the zero crossing of the $A_{FB}$ ($\mu$ mode) at $q^2 \simeq 1$ GeV$^2$ in the SM. In the presence of NP with tensor coupling, the zero crossing coincides with the SM value. On the other hand, for the $\tau$ mode the zero crossing occurs at $q^2 \simeq 14$ GeV$^2$ in the presence of NP without tensor operator.  However, when the tensor coupling is included, we see such  zero-crossing at $q^2 \simeq 11.2$ GeV$^2$.
\item ${\cal R}_{a_1}$ : We provide the prediction for the LFU violating observable ${\cal R}_{a_1}$, both in the SM and in the presence of new physics scenarios in Fig.~\ref{OBsbtoa1} as well as in Table  \ref{Tab:Numresults}. From  Table \ref{Tab:Numresults}, one can observe  a significant discrepancy between the SM and NP contributions.
\end{itemize}

\section{Conclusion}\label{conc}
Inspired by the anomalies present in the charged-current mediated $b \to c \ell \nu$ transitions,  we have studied the CKM suppressed semileptonic $B \to a_1 \ell \nu$ decay mode, which is  mediated through $b \to u \ell \nu$ transition, in the SMEFT framework. Considering  the WET Lagrangian describing the $b \to u \ell \nu$ transitions, we correlate the NP couplings in terms of the SMEFT couplings. We then constrained the parameter space for the new couplings by using the experimental available data of the branching fractions of $B \to \ell \bar{\nu}$ and $B \to (\pi, \rho, \omega) \ell \bar{\nu}$ processes in the presence of individual as well as mixed couplings of SMEFT Wilson coefficients. Using these constrained new couplings, we investigate the branching fractions, longitudinal and transverse angular coefficients as well as the forward-backward asymmetry parameters of $B \to a_1 \ell \nu$ processes for both muonic and tauonic final states for various NP scenarios.
We found that these observables show appreciable deviations from their SM predictions in the presence of NP operators as summarized in Table \ref{Tab:Numresults}. It should be emphasized that with the inclusion of $S+V$ type NP contributions, the branching fraction of $B \to a_1 \mu \nu$ is slightly reduced from its SM value while for $S+V+T$ type it remains consistent with the SM result. On the other hand, for tauonic mode the branching fraction deviated significantly from SM for both type of NP scenarios. The forward-backward asymmetry parameter remains more or less consistent with the SM result.  To summarize, we have performed a comprehensive study of $B \to a_1 \ell \nu$ decay mode and found that the branching fractions both in the SM as well as in the NP scenarios are within the sensitivity reach of currently running Belle-II and LHCb experiments.  The precise determination of this branching ratio would establish or rule out the possible role of new physics in $b \to u \ell \nu$ transition.

\acknowledgments 
MKM would like to acknowledge IoE PDRF, University of Hyderabad for the financial support. DP acknowledges the support of Prime Minister's Research Fellowship, Government of India. RM would like to acknowledge University of Hyderabad IoE project grant no. RC1-20-012.

\appendix 
\section{Hadronic Matrix Elements and the Helicity amplitudes}
The hadronic matrix elements for $B \to a_1$ transition are written in terms of the form factors as follows~\cite{Colangelo:2020jmb},
\bea
\langle a_1(p^\prime,\epsilon)|{\bar u} \gamma_\mu(1-\gamma_5) b| {\bar B}(p) \rangle &=&
 {2 A^{B \to a_1}(q^2) \over m_B+m_{a_1}} i \epsilon_{\mu \nu \alpha \beta} \epsilon^{*\nu}  p^\alpha p^{\prime \beta} \nn \\
&+&\Big\{ (m_B+m_{a_1}) \left[ \epsilon^*_\mu -{(\epsilon^* \cdot q) \over q^2} q_\mu \right] V_1^{B \to a_1}(q^2) \nn\\
&-& {(\epsilon^* \cdot q) \over  m_B+m_{a_1}} \left[ (p+p^\prime)_\mu -{m_B^2-m_{a_1}^2 \over q^2} q_\mu \right] V_2^{B \to a_1}(q^2) \nn \\
&+& (\epsilon^* \cdot q){2 m_{a_1} \over q^2} q_\mu V_0^{B \to a_1}(q^2) \Big\}  \label{FF-a1}
\eea
with the condition  $\displaystyle V_0^{B \to a_1}(0)= \frac{m_B + m_{a_1}}{2 m_{a_1}} V_1^{B \to a_1}(0) -  \frac{m_B - m_{a_1}}{2 m_{a_1}}  V_2^{B \to a_1}(0)$, and
\bea
\langle a_1(p^\prime,\epsilon)|{\bar u}  b| {\bar B}(p) \rangle &=&\frac{2 m_{a_1}}{m_b-m_u} (\epsilon^* \cdot q) V_0^{B \to a_1}(q^2)
\label{scalar-a1} \\
\langle  a_1(p^\prime,\epsilon)|{\bar u} \sigma_{\mu \nu}b| {\bar B}(p) \rangle &=&
i\, T_0^{B \to  a_1}(q^2) {\epsilon^* \cdot q \over (m_B+ m_{ a_1})^2} (p_\mu p^\prime_\nu-p_\nu p^\prime_\mu) \nn \\
&& +i\,
T_1^{B \to  a_1}(q^2) (p_\mu \epsilon_\nu^*-\epsilon_\mu^* p_\nu)+i\,T_2^{B \to  a_1}(q^2)(p^\prime_\mu \epsilon_\nu^*-\epsilon_\mu^* p^\prime_\nu)  \,\,\,\,\,\,\,  \label{mat-tensor-a1} \\
\langle a_1(p^\prime,\epsilon)|{\bar u} \sigma_{\mu \nu} \gamma_5  b| {\bar B}(p) \rangle& =&
T_0^{B \to a_1}(q^2) {\epsilon^* \cdot q \over (m_B+ m_ {a_1})^2} \epsilon_{\mu \nu \alpha \beta} p^\alpha p^{\prime \beta}\nn \\
&+&
T_1^{B \to  a_1}(q^2) \epsilon_{\mu \nu \alpha \beta} p^\alpha \epsilon^{*\beta}+ T_2^{B \to a_1}(q^2) \epsilon_{\mu \nu \alpha \beta} p^{\prime \alpha} \epsilon^{*\beta} \label{mat-tensor1-a1}
\eea
The helicity amplitudes required for the $B \to a_1 \ell \nu$ process are given as follows,
\bea
H_0^{a_1} &=&\frac{-(m_B+m_{a_1} )^2(m_B^2-m_{a_1} ^2-q^2) V_1(q^2)+\lambda(m_B^2,\,m_{a_1} ^2,\,q^2) V_2(q^2)}{2m_{a_1}(m_B+m_{a_1} ) \sqrt{q^2}} \nn \\
H_\pm^{a_1} &=& \frac{-(m_B+m_{a_1} )^2 V_1(q^2)\pm\sqrt{\lambda(m_B^2,\,m_{a_1} ^2,\,q^2)}A(q^2)}{m_B+m_{a_1} }  \label{Hampa1}\\
H_t^{a_1} &=& \frac{\sqrt{\lambda(m_B^2,\,m_{a_1} ^2,\,q^2)}}{\sqrt{q^2}} \,V_0(q^2) \,\,\, . \nn
\eea
For the  tensor operator, the associated helicity amplitudes are given as \cite{Colangelo:2018cnj}:
\bea
H_+^{NP} &=& \frac{1}{\sqrt{q^2}}\left\{\left[m_B^2-m_{a_1}^2+\lambda^{1/2} (m_B^2,m_{a_1}^2,q^2) \right](T_1^{B \to a_1}+ T_2^{B \to a_1})+q^2(T_1^{B \to a_1}- T_2^{B \to a_1})\right\} \nn \\
H_-^{NP} &=& \frac{1}{\sqrt{q^2}}\left\{\left[m_B^2-m_{a_1}^2-\lambda^{1/2} (m_B^2,m_{a_1}^2,q^2) \right](T_1^{B \to a_1}+ T_2^{B \to a_1})+q^2(T_1^{B \to a_1}-T_2^{B \to a_1})\right\} \hspace*{1cm} \\
H_L^{NP}&=&4\Big\{
\frac{\lambda (m_B^2,m_{a_1}^2,q^2)}{m_{a_1}(m_B+m_{a_1})^2} \, T_0^{B \to a_1}+2\frac{m_B^2+m_{a_1}^2-q^2}{m_{a_1}}\, T_1^{B \to a_1}
+4m_{a_1}\, T_2^{B \to a_1} \Big\} \,\, .\nn
\eea
\section{Angular coefficient functions $I_i^{SM, NP}$ (Longitudinal and Transverse)}\label{app:coeff}
The details of the angular coefficients in the SM are given as follows:
\begin{table}[htb]
\scalebox{0.85}{
\begin{tabular}{ccc}
\hline
\hline 
\noalign{\medskip}
$i$ & $I_{i,||}$ & $I_{i,\perp}$ \\
\noalign{\medskip}
\hline
\noalign{\medskip}
$I_{1s}^{a_1}$ & $\frac{1}{2}(H_+^2 + H_-^2)(m_{\ell}^2 + 3 q^2)$ & $2 H_t^2 m_\ell^2 + H_0^2 (m_\ell^2 + q^2) + \frac{1}{4} (H_+^2 + H_-^2)(m_{\ell}^2 + 3 q^2)$ \\
\hline
\noalign{\medskip}
$I_{1c}^{a_1}$ & $4 H_t^2 m_{\ell}^2 + 2 H_0^2 (m_{\ell}^2 + q^2)$ & $\frac{1}{2} (H_+^2 + H_-^2)(m_{\ell}^2 + 3 q^2)$ \\
\hline
\noalign{\medskip}
$I_{2s}^{a_1}$ & $- \frac{1}{2}(H_+^2 + H_-^2)(m_{\ell}^2 - q^2)$ & $[H_0^2 - \frac{1}{4}(H_+^2 + H_-^2)](m_{\ell}^2 - q^2)$ \\
\hline
\noalign{\medskip}
$I_{2c}^{a_1}$ & $2 H_0^2 (m_{\ell}^2 - q^2)$ & $- \frac{1}{2} (H_+^2 + H_-^2)(m_{\ell}^2 - q^2)$ \\
\hline
\noalign{\medskip}
$I_{3}^{a_1}$ & $2 H_+ H_- (m_{\ell}^2 - q^2)$ & $- H_+ H_- (m_{\ell}^2 - q^2)$ \\
\hline
\noalign{\medskip}
$I_{4}^{a_1}$ & $H_0 (H_+ + H_-) (m_{\ell}^2 - q^2)$ & $- \frac{1}{2} H_0 (H_+ + H_-)(m_{\ell}^2 - q^2)$ \\
\hline
\noalign{\medskip}
$I_{5}^{a_1}$ & $-2 H_t (H_+ + H_-) m_{\ell}^2 - 2 H_0 (H_+ - H_-) q^2$ & $H_t (H_+ + H_-) m_\ell^2 + H_0 (H_+ - H_-) q^2$ \\
\hline
\noalign{\medskip}
$I_{6s}^{a_1}$ & $2 (H_+^2 - H_-^2) q^2$ & $-4 H_t H_0 m_\ell^2 + (H_+^2 - H_-^2) q^2$ \\
\hline
\noalign{\medskip}
$I_{6c}^{a_1}$ & $- 8 H_t H_0 m_{\ell}^2$ & $2 (H_+^2 - H_-^2) q^2$ \\
\hline
\noalign{\medskip}
$I_{7}^{a_1}$ & $0$ & $0$ \\
\noalign{\medskip}
\hline
\hline
\end{tabular}}
\caption{\small Angular coefficient functions  in the four dimensional $\bar B \to a_1 (\rho \pi) \ell^- \bar \nu_\ell$ decay distribution given in  Eq.\ref{angulara1}, in SM.}
\label{tab:a1SM}
\end{table}
\newpage
The details of the angular coefficients in the new physics are given as follows:
\begin{table}[htp]
\caption{ \small Angular coefficient functions for   $\bar B \to a_1 (\rho \pi)  \ell^- \bar \nu_\ell$: NP term with S operator,  interference  SM-NP with S operator, and   NP-NP interference with S and T operators, Eq.\eqref{eq:Iang}. }\label{tab:a1parS}
\scalebox{0.84}{
\begin{tabular}{cccc}
\hline
\hline
\noalign{\medskip}
$i$ & $I_{i,||}^{NP, S}$ & $I_{i,||}^{INT, S}$ & $I_{i,||}^{INT, ST}$\\
\noalign{\medskip}
\hline
\noalign{\medskip}
$I_{1s}^{a_1}$ & $0$ & $0$ & $0$ \\
\noalign{\medskip}
$I_{1c}^{a_1}$ & $4 H_t^2 \frac{q^4}{(m_b - m_u)^2}$ & $4 H_t^2 \frac{m_\ell q^2}{m_b-m_u}$ &$0$ \\
\hline
\noalign{\medskip}
$I_{2s}^{a_1}$ & $0$ & $0$ &$0$ \\
\hline
\noalign{\medskip}
$I_{2c}^{a_1}$ & $0$ & $0$ &$0$\\
\hline
\noalign{\medskip}
$I_{3}^{a_1}$ & $0$ & $0$ &$0$\\
\hline
\noalign{\medskip}
$I_{4}^{a_1}$ & $0$ & $0$ &$0$\\
\hline
\noalign{\medskip}
$I_{5}^{a_1}$ & $0$ & $- H_t (H_+ + H_-) \frac{m_\ell q^2}{m_b-m_u}$ & $- 2H_t(H_+^{NP} + H_-^{NP}) \frac{ (q^2)^{3/2}}{m_b - m_u}$ \\
\hline
\noalign{\medskip}
$I_{6s}^{a_1}$ & $0$ & $0$ &$0$\\
\hline
\noalign{\medskip}
$I_{6c}^{a_1}$ & $0$ & $-4 H_t H_0 \frac{m_\ell q^2}{m_b-m_u}$ & \,\,$- H_t\, H_L^{NP} \frac{ (q^2)^{3/2}}{m_b - m_u}$\\
\hline
\noalign{\medskip}
$I_{7}^{a_1}$ & $0$ & $- H_t (H_+ - H_-) \frac{m_\ell q^2}{m_b-m_u}$ & \,\, $ -2H_t(H_+^{NP} - H_-^{NP}) \frac{ (q^2)^{3/2}}{m_b - m_u}$ \\
\noalign{\medskip}
\hline
\hline
\end{tabular}} 
\end{table}
\begin{table}[htp] 
\scalebox{0.865}{
\begin{tabular}{cccc}
\hline
\hline
\noalign{\medskip}
$i$ & $I_{i,\perp}^{NP, S}$ & $I_{i,\perp}^{INT, S}$ & $I_{i,\perp}^{INT,ST}$\\
\noalign{\medskip}
\hline
\noalign{\medskip}
$I_{1s}^{a_1}$ & $2 H_t^2 \frac{q^4}{(m_b - m_u)^2}$ & $2 H_t^2 \frac{m_\ell q^2}{m_b-m_u}$&$0$ \\
\hline
\noalign{\medskip}
$I_{1c}^{a_1}$ & $0$ & $0$ & $0$ \\
\hline
\noalign{\medskip}
$I_{2s}^{a_1}$ & $0$ & $0$ & $0$ \\
\hline
\noalign{\medskip}
$I_{2c}^{a_1}$ & $0$ & $0$ & $0$ \\
\hline
\noalign{\medskip}
$I_{3}^{a_1}$ & $0$ & $0$ & $0$  \\
\hline
\noalign{\medskip}
$I_{4}^{a_1}$ & $0$ & $0$ & $0$ \\
\hline
\noalign{\medskip}
$I_{5}^{a_1}$ & $0$ & $\frac{1}{2} H_t (H_+ + H_-) \frac{m_\ell q^2}{m_b-m_u}$ & $ H_t(H_+^{NP} + H_-^{NP}) \frac{ (q^2)^{3/2}}{m_b - m_u}$ \\
\hline
\noalign{\medskip}
$I_{6s}^{a_1}$ & $0$ & $-2 H_t H_0 \frac{m_\ell q^2}{m_b-m_u}$ &  \,\,$- H_t\, H_L^{NP} \frac{ (q^2)^{3/2}}{2(m_b - m_u)}$ \\
\hline
\noalign{\medskip}
$I_{6c}^{a_1}$ & $0$ & $0$ & $0$ \\
\hline
\noalign{\medskip}
$I_{7}^{a_1}$ & $0$ & $\frac{1}{2} H_t (H_+ - H_-) \frac{m_\ell q^2}{m_b-m_u}$ & \,\, $ H_t(H_+^{NP} - H_-^{NP}) \frac{ (q^2)^{3/2}}{m_b - m_u}$ \\
\noalign{\medskip}
\hline
\hline
\end{tabular}}
\caption{ \small Similar as Table \ref{tab:a1parS} with transverse coefficients. }\label{tab:a1perpS}
\end{table}
\small
\begin{table}[htp]
\centering
\scalebox{0.7}{
\begin{tabular}{ccc}
\hline
\hline
\noalign{\medskip}
$i$ & $I_{i,||}^{NP, T}$ & $I_{i,||}^{INT, T}$ \\
\noalign{\medskip}
\hline
\noalign{\medskip}
$I_{1s}^{a_1}$ & $2[(H_+^{NP})^2 + (H_-^{NP})^2] (3 m_\ell^2 + q^2)$ & $4 ( H_+^{NP} H_+ + H_-^{NP} H_- ) m_\ell \sqrt{q^2}$ \\
\hline
\noalign{\medskip}
$I_{1c}^{a_1}$ & $\frac{1}{8} (H_L^{NP})^2 (m_\ell^2 + q^2)$ & $ H_L^{NP} H_0 m_\ell \sqrt{q^2}$ \\
\hline
\noalign{\medskip}
$I_{2s}^{a_1}$ & $2[(H_+^{NP})^2+(H_-^{NP})^2](m_\ell^2 - q^2)$ & $0$ \\
\hline
\noalign{\medskip}
$I_{2c}^{a_1}$ & $- \frac{1}{8} (H_L^{NP})^2 (m_\ell^2 - q^2)$ & $0$ \\
\hline
\noalign{\medskip}
$I_{3}^{a_1}$ & $-8 H_+^{NP} H_-^{NP} (m_\ell^2 - q^2)$ & $0$ \\
\hline
\noalign{\medskip}
$I_{4}^{a_1}$ & $- \frac{1}{2} H_L^{NP} (H_+^{NP} + H_-^{NP}) (m_\ell^2 - q^2)$ & $0$ \\
\hline
\noalign{\medskip}
$I_{5}^{a_1}$ & $- H_L^{NP} (H_+^{NP} - H_-^{NP}) m_\ell^2$ & $-\frac{1}{4} [H_L^{NP} (H_+ - H_-) + 8 H_+^{NP} (H_t + H_0)  $ \\
\hline
\noalign{\smallskip}
$\quad$ & $\quad$ & $ \qquad \qquad + 8 H_-^{NP} (H_t - H_0)] m_\ell \sqrt{q^2}$ \\
\hline
\noalign{\medskip}
$I_{6s}^{a_1}$ & $8 [(H_+^{NP})^2 - (H_-^{NP})^2] m_\ell^2$ & $4 ( H_+^{NP} H_+ - H_-^{NP} H_- ) m_\ell \sqrt{q^2}$ \\
\hline
\noalign{\medskip}
$I_{6c}^{a_1}$ & $0$ & $-H_L^{NP} H_t m_{\ell} \sqrt{q^2}$ \\
\hline
\noalign{\medskip}
$I_{7}^{a_1}$ & $0$ & $-\frac{1}{4} [H_L^{NP} (H_+ + H_-) - 8 H_+^{NP} (H_t + H_0)  $ \\
\hline
\noalign{\smallskip}
$\quad$ & $\quad$ & $\qquad \qquad + 8 H_-^{NP} (H_t - H_0)] m_\ell \sqrt{q^2}$ \\
\noalign{\medskip}
\hline
\hline
\end{tabular}}
\caption{ \small Angular coefficient functions for  $\bar B \to a_1 (\rho \pi)  \ell^- \bar \nu_\ell$:  NP term with T operator and interference  SM-NP with T operator. }\label{tab:a1parT}
\end{table}
\begin{table}[htp]
\centering
\scalebox{0.7}{
\begin{tabular}{ccc}
\hline
\hline
\noalign{\medskip}
$i$ & $I_{i,\perp}^{{NP}, T}$ & $I_{i,\perp}^{INT, T}$ \\
\noalign{\medskip}
\hline
\noalign{\medskip}
$I_{1s}^{a_1}$ & $[(H_+^{NP})^2 + (H_-^{NP})^2] (3 m_\ell^2 + q^2)$ & $\frac{1}{2} [4 (H_+^{NP} H_+ + H_-^{NP} H_- )  $ \\
\hline
\noalign{\smallskip}
$\quad$ & $\qquad \qquad \quad + \frac{1}{16} (H_L^{NP})^2 (m_\ell^2 + q^2) $ & $\qquad \qquad + H_L^{NP} H_0] m_\ell \sqrt{q^2}$ \\
\hline
\noalign{\medskip}
$I_{1c}^{a_1}$ & $2 [(H_+^{NP})^2 + (H_-^{NP})^2] (3m_\ell^2 + q^2)$ & $ 4 (H_+^{NP} H_+ + H_-^{NP} H_- ) m_\ell \sqrt{q^2}$ \\
\hline
\noalign{\medskip}
$I_{2s}^{a_1}$ & $[(H_+^{NP})^2 + (H_-^{NP})^2] (m_\ell^2 - q^2)  $ & $0$ \\
\hline
\noalign{\smallskip}
$\quad$ & $\qquad \qquad \quad - \frac{1}{16} (H_L^{NP})^2 (m_\ell^2 - q^2)  $ & $\quad$ \\
\hline
\noalign{\medskip}
$I_{2c}^{a_1}$ & $2 [(H_+^{NP})^2 + (H_-^{NP})^2] (m_\ell^2 - q^2)$ & $0$ \\
\hline
\noalign{\medskip}
$I_{3}^{a_1}$ & $4 H_+^{NP} H_-^{NP} (m_\ell^2 - q^2)$ & $0$ \\
\hline
\noalign{\medskip}
$I_{4}^{a_1}$ & $\frac{1}{4} H_L^{NP} (H_+^{NP} + H_-^{NP}) (m_\ell^2 - q^2)$ & $0$ \\
\hline
\noalign{\medskip}
$I_{5}^{a_1}$ & $\frac{1}{2} H_L^{NP} (H_+^{NP} - H_-^{NP}) m_\ell^2$ & $ \frac{1}{8} [H_L^{NP} (H_+ - H_-) + 8 H_+^{NP} (H_t + H_0)  $ \\
\hline
\noalign{\smallskip}
$\quad$ & $\quad$ & $\qquad \qquad + 8H_-^{NP} (H_t - H_0)] m_\ell \sqrt{q^2}$ \\
\hline
\noalign{\medskip}
$I_{6s}^{a_1}$ & $4 [(H_+^{NP})^2 - (H_-^{NP})^2] m_\ell^2$ & $-\frac{1}{2} [ -4 ( H_+^{NP} H_+ - H_-^{NP} H_- ) + H_L^{NP} H_t] m_\ell \sqrt{q^2}$ \\
\hline
\noalign{\medskip}
$I_{6c}^{a_1}$ & $8 [(H_+^{NP})^2 - (H_-^{NP})^2] m_\ell^2$ & $4 (H_+^{NP} H_+ - H_-^{NP} H_-) m_{\ell} \sqrt{q^2}$ \\
\hline
\noalign{\medskip}
$I_{7}^{a_1}$ & $0$ & $ \frac{1}{8} [H_L^{NP} (H_+ + H_-) - 8 H_+^{NP} (H_t + H_0)  $ \\
\hline
\noalign{\smallskip}
$\quad$ & $\quad$ & $\qquad \qquad + 8 H_-^{NP} (H_t - H_0)]$ \\
\noalign{\medskip}
\hline
\hline
\end{tabular}}
\caption{Similar as Table \ref{tab:a1parT} with transverse coefficients.}\label{tab:a1perpT}
\end{table}

\newpage

\bibliographystyle{ieeetr}
\bibliography{MSDR}

\end{document}